\documentclass[useAMS,usenatbib]{mn2e}

\usepackage{graphicx,color}
\usepackage{amsmath}
\usepackage{amssymb}
\usepackage{bm}
\usepackage{ulem}
\usepackage{enumerate}
\usepackage{extarrows}

\defcitealias{Chamandy+13a}{CSS13a}
\defcitealias{Chamandy+13b}{CSS13b}
\defcitealias{Comparetta+Quillen12}{CQ12}

\definecolor{webgreen}{rgb}{0,.5,0}
\definecolor{webbrown}{rgb}{.6,0,0}
\newcommand\deriv[2]{\displaystyle\frac{\partial #1}{\partial #2} }

\newcommand{\What}{\widehat{W}}
\newcommand{\Vhat}{\widehat{V}}
\newcommand{\Exp}[1]{{\rm e}^{#1}}

\newcommand{\del}{\partial}
\newcommand{\Del}{{\nabla}}

\newcommand{\bmDel}{\bm{\nabla}}

\newcommand{\alp}{\alpha}

\newcommand{\Emf}{\bm{\mathcal{E}}}

\newcommand{\Flux}{\bm{\mathcal{F}}}
\newcommand{\Bs}{\mathcal{B}}
\newcommand{\Bsr}{\mathcal{B}_r}
\newcommand{\Bsp}{\mathcal{B}_\phi}

\newcommand{\bfu}{\bm{u}}
\newcommand{\bfb}{\bm{b}}
\newcommand{\bmb}{\bm{b}}

\newcommand{\mean}[1]{\overline{#1}}
\newcommand{\meanv}[1]{\overline{\bm{#1}}}

\newcommand{\D}{_\mathrm{D}}						%disc
\newcommand{\eq}{_\mathrm{eq}}						%equipartition
\newcommand{\f}{_\mathrm{0}}					   	%forcing
\newcommand{\kin}{_\mathrm{k}}			   		%kinematic
\newcommand{\magn}{_\mathrm{m}}			   		%magnetic
\newcommand{\turb}{_\mathrm{t}}			   		%turbulent
\newcommand{\etat}{\eta_\mathrm{t}}			   	%turbulent diffusivity
\newcommand{\crit}{_\mathrm{c}}			   		%critical
\newcommand{\const}{\mathrm{const}}			   	%constant
\newcommand{\dd}{\mathrm{d}}			   		%differential
\newcommand{\adv}{_\mathrm{a}}			   		%advection, subscript
\newcommand{\diff}{_\mathrm{d}}			   		%diffusion, subscript
\newcommand{\Adv}{\mathrm{a}}			   		%advection, subscript
\newcommand{\Diff}{\mathrm{d}}			   		%diffusion, in-line

\newcommand{\cro}{\times}
\newcommand{\Rm}{\mathcal{R}_\mathrm{m}}
\newcommand{\mbr}{\mean{B}_r}
\newcommand{\mbp}{\mean{B}_\phi}
\newcommand{\mbz}{\mean{B}_z}

\newcommand{\mup}{\mean{U}_\phi}
\newcommand{\muz}{\mean{U}_z}

\newcommand{\alphatilde}{\widetilde{\alpha}}

  \newcommand{\km}{\,{\rm km}}
  \newcommand{\kms}{\,{\rm km\,s^{-1}}}
  \newcommand{\kmskpc}{\,{\rm km\,s^{-1}\,kpc^{-1}}}
  \newcommand{\cmcms}{\,{\rm cm^2\,s^{-1}}}

  \newcommand{\kpc}{\,{\rm kpc}}
  \newcommand{\pc}{\,{\rm pc}}
  
  \newcommand{\Myr}{\,{\rm Myr}}
  
  \newcommand{\Gyr}{\,{\rm Gyr}}

  \newcommand{\mkG}{\,\mu{\rm G}}

  \newcommand{\s}{\,{\rm s}}

\pagerange{\pageref{firstpage}--\pageref{lastpage}} 
\pubyear{2014}    
\label{firstpage}

\begin{document}

\title[Non-linear galactic dynamos: A toolbox]{Non-linear galactic dynamos: A toolbox}
\author[L.\ Chamandy et al.]{Luke Chamandy,$^{1}$\thanks{E-mail: luke@iucaa.ernet.in;
																			anvar.shukurov@ncl.ac.uk;
																			kandu@iucaa.ernet.in; 
																			k.j.stoker@googlemail.com} 
														Anvar Shukurov,$^{2,1}$
														Kandaswamy Subramanian,$^{1}$
\newauthor % starts a new line in the author environment 
														Katherine Stoker$^{2}$\\
$^{1}$Inter-University Centre for Astronomy and Astrophysics, Post Bag 4, Ganeshkhind, Pune 411007, India\\
$^{2}$School of Mathematics \& Statistics, Newcastle University, Newcastle upon Tyne NE1 7RU}

%\date{Submitted to MNRAS}

\maketitle

%-------------------------------------------------------------------------
\begin{abstract}
We compare various models and approximations for non-linear mean-field dynamos 
in disc galaxies to assess their applicability and accuracy, 
and thus to suggest a set of simple solutions suitable to model the large-scale galactic magnetic fields in various contexts. 
The dynamo saturation mechanisms considered are the magnetic helicity balance involving helicity fluxes 
(the dynamical $\alpha$-quenching) and an algebraic $\alpha$-quenching. 
The non-linear solutions are then compared with the marginal kinematic and asymptotic solutions. 
We also discuss the accuracy of the no-$z$ approximation.
Although these tools are very different in the degree of approximation and hence complexity, 
they all lead to remarkably similar solutions for the mean magnetic field. 
In particular, we show that the algebraic $\alpha$-quenching non-linearity
can be obtained from a more physical dynamical $\alpha$-quenching model
in the limit of nearly azimuthal magnetic field. 
This suggests, for instance, 
that earlier results on galactic disc dynamos based on the simple algebraic non-linearity are likely to be reliable, 
and that estimates based on simple, even linear models are often a good starting point. 
We suggest improved no-$z$ and algebraic $\alpha$-quenching models, 
and also incorporate galactic outflows into a simple analytical dynamo model to show 
that the outflow can produce leading magnetic spirals near the disc surface. 
The simple dynamo models developed are applied to estimate the magnetic pitch angle 
and the arm-interarm contrast in the saturated magnetic field strength for realistic parameter values.
\end{abstract}


%----------------------------------------------------------------------------
\section{Introduction}
The mean-field dynamo theory provides an appealing explanation
of the presence and structure of large-scale magnetic fields in disc galaxies 
\citep{Ruzmaikin+88,Beck+96,Brandenburg+Subramanian05a,Shukurov05,Kulsrud+Zweibel08}.
The dynamo time scale is shorter than the galactic lifetime, and the energy densities
of the large-scale galactic magnetic fields and interstellar turbulence are observed 
to be of the same order of magnitude.
It is thus plausible that the galactic large-scale dynamos are normally in a 
non-linear, statistically steady state. Recent progress in dynamo
theory has lead to physically motivated non-linear models 
where the steady state is achieved through the magnetic helicity balance
in the dynamo system \citep[reviewed  by][]{Brandenburg+Subramanian05a,Blackman14}. 
To avoid a catastrophic suppression of the mean induction effects of turbulence, 
the magnetic helicity of random magnetic fields should be removed from 
the system. In galaxies, this can be achieved through the advection of magnetic fields
from the disc to the halo by the galactic fountain or wind \citep{Shukurov+06,Sur+07},
diffusive flux \citep{Kleeorin+00,Kleeorin+02} and helicity flux relying on
the anisotropy of the interstellar turbulence \citep{Vishniac+Cho01,Vishniac+Shapovalov14} 
\citep[see ][for an application to galaxies]{Sur+07}. The first of these mechanisms 
is the simplest in physical and mathematical terms, and involves galactic parameters 
that are reasonably well constrained observationally.

Most of the earlier analytical and numerical results in the non-linear mean-field disc dynamo theory 
rely on a much simpler form of non-linearity in the dynamo equations, the so-called
algebraic $\alpha$-quenching that is based on a simple, explicit form of the dependence of the 
dynamo parameters, usually the $\alpha$-coefficient, on the magnetic field.
In a thin layer, such as a galactic or accretion disc, this allows one to obtain a wide range of analytical
and straightforward numerical solutions using simple and yet accurate approximations 
\citep[e.g.,][]{Shukurov04}.
One of the advantages of the resulting theory of galactic magnetic fields is that all
its essential parameters can be expressed in terms of observable quantities (the 
angular velocity of rotation, thickness of the gas layer, turbulent velocity, etc.).
As a result, theory of galactic magnetic fields has been better constrained and verified by 
direct comparison with observations than, arguably, any other astrophysical dynamo theory.
Such comparisons require relatively simple, preferably analytical, approximations to the
solutions of the dynamo equations. In this paper we consider numerical solutions of 
thin-disc dynamo equations with a dynamic non-linearity involving magnetic helicity balance
and compare them with a wide range of simpler solutions to develop a set of accessible
tools to facilitate applications of the theory.

The paper is organized as follows.
In Sections~\ref{sec:dynamo}-\ref{sec:equations}, we present the theoretical background and
a review of each of the approximations discussed.
This is followed by a detailed comparison of the solutions resulting from
various physical and mathematical approximations in Section \ref{sec:results}.
In particular, in Section~\ref{sec:alg} we provide an in-depth comparison of 
the dynamical and algebraic non-linearities. Our overall conclusion is that
the earlier, simple models, when applied judiciously, reproduce comfortably
well solutions with the dynamical non-linearity.
Section~\ref{sec:examples} provides examples of
applications of the toolbox, 
namely the magnetic pitch angle problem and the  
spiral arm-interarm contrasts in magnetic field.
We present a summary and general conclusions in Section \ref{sec:conc}.
The details of the asymptototic solutions studied,
namely the perturbation and no-$z$ solutions,
are given in Appendices~\ref{sec:pert_app} and \ref{sec:noz_app}, respectively.

Throughout this paper, we use cylindrical polar coordinates $(r,\phi,z)$ with the origin at the
disc centre and the $z$-axis aligned with the angular velocity of rotation $\bm\Omega$.

%------------------------------------------------------------------------
\section{Nonlinear mean-field dynamos}
\label{sec:dynamo}
%-------------------------------------------------------------------------
Magnetic field averaged over scales exceeding the turbulent scales, the mean magnetic
field $\meanv{B}$, is governed by the induction equation suitably averaged
in a mirror-asymmetric random flow \citep{Moffatt78, Krause+Radler80}:
\begin{equation}
  \label{dynamo}
  \frac{\del\meanv{B}}{\del t} =\Del\cro\left(\meanv{U}\cro\meanv{B}+\alpha\meanv{B}
  	-\eta\turb\Del\cro\meanv{B}\right)\,
\end{equation}
where overbar denotes averaged quantities, 
$\eta\turb$ is the magnetic diffusivity (dominated by the turbulent diffusion),
$\bm{B}$ is the magnetic field and $\bm{U}$ is the velocity field, 
both assumed to be separable into the mean and random parts,
\begin{equation*}
  \label{scaleseparation}
  \bm{U}= \meanv{U} +\bm{u}, \quad \bm{B}= \meanv{B} +\bm{b}\,,
\end{equation*}
with $\mean{\bm u}=\bm 0$ and $\mean{\bm b}=\bm 0$.
 
Following \citet{Pouquet+76,Kleeorin+95,Blackman+Field00}, the $\alpha$-effect in 
Eq.~\eqref{dynamo} is represented as the sum of kinetic and magnetic contributions,
\begin{equation*}
  \label{alp_total}
  \alpha= \alpha\kin +\alpha\magn,
\end{equation*}
where $\alpha\kin= -\tau\mean{{\bm u}\cdot\bmDel\cro{\bm u}}/3$ is 
the `kinetic' part related to the mean helicity of the random flow 
$\mean{{\bm u}\cdot\bmDel\cro{\bm u}}$, 
and $\alpha\magn= \tau\mean{\bmb\cdot\bmDel\cro\bmb}/(12\pi\rho)$
is the magnetic contribution (here $\rho$ is the gas density and $\tau$ is the correlation time of the random flow). 
The latter is responsible for non-linear dynamo effects: 
$\alpha\magn$ and $\alpha\kin$ usually have opposite signs and, as $\mean{\bm{b}\cdot \bmDel\cro\bm{b}}$ is amplified
(together with or independently of $\mean{\bm B}$), the magnitude of the total $\alpha$ effect
decreases, and this saturates the growth of the mean magnetic field.

The dynamo non-linearity resulting from the magnetic helicity conservation is
governed by the following equation for the magnetic contribution to the $\alpha$ effect:
\citep{Kleeorin+Ruzmaikin82,Kleeorin+95,Subramanian+Brandenburg06} \citep[see also]
[for a form adapted to disc dynamos]{Shukurov+06}:
\begin{equation}
  \label{dalpha_mdt}
  \frac{\del\alpha\magn}{\del t}
  =-\frac{2\eta\turb}{l^2B\eq^2}\Emf\cdot\meanv{B}-\bmDel\cdot\Flux,
\end{equation}
where $l$ is the outer scale of the turbulence, $B\eq=u\sqrt{4\pi\rho}$, with $\rho$ the gas density, 
is the characteristic strength of magnetic field, here taken to correspond to energy
equipartition with turbulence,
$\Emf=\mean{\bfu\cro\bfb}$ is the mean turbulent electromotive force,
and $\Flux$ is a flux density of $\alpha\magn$ (related to the flux density of the small-scale magnetic helicity density).
Ohmic dissipation has been neglected in equation \eqref{dalpha_mdt}, 
which is justified if the time scales considered are short compared to the resistive time scale
(exceeding the galactic lifetime at galactic scales) 
or if the Ohmic diffusion is negligible compared to the helicity flux.
In fact, the latter condition must hold in order to avert the catastrophic quenching of the dynamo.
The flux density can be written as the sum of several contributions,
\begin{equation*}
  \Flux=\Flux_\Adv +\Flux_\Diff + ... .
\end{equation*}
where the advective flux density $\Flux_\Adv$ and diffusive flux density $\Flux_\Diff$ 
are the two that are considered in this work,
although other contributions exist \citep{Vishniac+Cho01,Ebrahimi+Bhattacharjee14},
and, generally, can be stronger than the diffusive flux density \citep{Vishniac+Shapovalov14}.

The advective flux density is given by
\begin{equation*}
  \label{advectiveflux}
  \Flux_\Adv=\meanv{U}\alpha\magn.
\end{equation*}
A vertical advective flux of $\alpha\magn$ 
is expected to be present because of galactic winds and fountain flow
\citep{Shukurov+06,Heald12,Bernet+13}.
The diffusive flux density \citep{Kleeorin+02}
\begin{equation*}
  \label{diffusiveflux}
  \Flux_\Diff=-\kappa\bmDel\alpha\magn, 
\end{equation*}
has been detected in direct numerical simulations \citep{Brandenburg+09},
with $\kappa\approx0.3\eta\turb$ as obtained from numerical simulations \citep{Mitra+10}.

The idea and equations of the dynamic non-linearity due to the
magnetic helicity conservation were suggested in the early 1980s \citep{Kleeorin+Ruzmaikin82}. Nevertheless,
for 10--20 more years, simple, heuristic prescriptions
of the dependence of $\alpha$ on the mean magnetic field had been used widely until the 
essential role of the magnetic helicity balance in mean-field dynamos was fully appreciated
in response to the discovery of effects of the fluctuation dynamo on the non-linear states
of the mean-field dynamo \citep{Vainshtein+Cattaneo92}. 
Most popular was an algebraic form
\begin{equation}
  \label{algq}
  \alpha=\frac{\alpha\kin}{1+aB^2/B\eq^2}\,
\end{equation}
where $B\equiv|\meanv{B}|$, 
known as an algebraic $\alpha$-quenching prescription.
Here $a$ is a parameter of order unity which will be adjusted in Section~\ref{sec:alg_quench} so %what follows 
as to achieve agreement with results obtained with the dynamical non-linearity \eqref{dalpha_mdt}.
Until then, we assume $a=1$, as is standard.

For the sake of simplicity, $\alpha$ and $\eta\turb$ are here assumed to be 
pseudoscalar and scalar (as appropriate in isotropic turbulence). 
In the approximation of the $\alpha\omega$-dynamo,
the induction effects of the galactic differential rotation, which produces $\mean{B}_\phi$
from $\mean{B}_r$, are assumed to be stronger 
than the similar mean induction effects of the random flow. Then the relevant component of the $\alpha$-tensor
responsible for the generation of $\mean{B}_r$ from $\mean{B}_\phi$ is $\alpha_{\phi\phi}$.

%------------------------------------------------------------
\section{Galactic dynamos}
\label{sec:galaxy}
In a thin disc, the kinetic contribution to the $\alpha$-effect can be written as the product of 
$r$-dependent and $z$-dependent parts,
\begin{equation}
  \label{alp_k}
  \alpha\kin=\alpha\f(r)\,\alphatilde(z)\,,
\end{equation}
where $\alpha\f$ can be estimated as 
\citep{Krause+Radler80,Ruzmaikin+88,Brandenburg+Subramanian05a},
\begin{equation}
  \label{Krause}
  \alpha\f=\frac{l^2\Omega}{h}\,,
\end{equation}
with $h$ the disc half-thickness, a function of $r$ in a flared disc, $\Omega$
the angular velocity of the gas, also a function of $r$, and $l$ the correlation
scale of the random velocity field. It follows from symmetry considerations
that $\alpha$ is an odd function of $z$ and $\alpha>0$ for $z>0$ \citep{Ruzmaikin+88}. 
As in numerous earlier analytical studies of the mean-field disc dynamos, 
we adopt
\[
  \alphatilde= \sin\left(\frac{\pi z}{h}\right)\,.
\]

The mean velocity in a disc galaxy is dominated by the azimuthal component, representing
differential rotation, and the vertical (outflow) velocity, a wind or a fountain flow:
\[
\meanv{U}= (0,r\Omega,\muz)\,,
\]
where $\muz$ is the mass-weighted outflow speed.
At small distances from the midplane, one can use
\begin{equation}\label{muztilde}
  \muz= U\f \widetilde{U}_z\,,
  \qquad
  \widetilde{U}_z=\frac{z}{h}\,.  
\end{equation}

We use the axisymmetric disc model of \citet{Chamandy+13a}
that has a thin, stratified, differentially rotating, flared, 
turbulent disc with the turbulent scale and rms velocity 
of $l=0.1\kpc$ and $u=10\kms$, and the mixing-length estimate of the turbulent 
magnetic diffusivity follows as
\begin{equation}
  \label{etat}
  \etat=\tfrac{1}{3}lu=10^{26}\cmcms.
\end{equation}
We also adopt $U\f=1\kms$, constant with radius \citep{Sur+07}.
For the galactic rotation curve, the Brandt's form
\begin{equation*}
  \label{Brandt}
  \Omega(r)=\frac{\Omega_0}{\left[1+(r/r_\omega)^2\right]^{1/2}}\,,
\end{equation*}
with $r_\omega=2\kpc$, and $U_\phi= 250\kmskpc$ at $r=10\kpc$ is chosen,
resulting in $\Omega_0\simeq127\kmskpc$.

The disc half-thickness is assumed to vary hyperbolically with radius,
$h(r)=h\D\left[1+(r/r\D)^2\right]^{1/2}$,
where $h\D$ is the scale height at $r=0$, and $r\D=10\kpc$ controls the disc flaring rate.
The value of $h$ at $r=10\kpc$ is chosen to be $0.5\kpc$, which gives $h\D\simeq0.35\kpc$.
The equipartition magnetic field strength is taken as
\begin{equation*}
  B\eq=B\f\exp{\left[-\frac{r}{R}-\frac{z^2}{2h^2}\right]},
\end{equation*}
where $R=20\kpc$.
This is equivalent to an exponential scale length of $10\kpc$ for the turbulent energy in the ionized gas.
For comparison, this is similar to but slightly larger than the scale length $\sim7\kpc$ 
of the total magnetic energy in the galaxy NGC~6946 \citep{Beck07}.
The value of $B\f$ can be chosen as convenient, 
e.g. $B\f=8.2\mkG$ to have $B\eq\simeq5\mkG$ at $r=10\kpc$, $z=0$.

We consider in detail specific models at $r=4\kpc$ and $r=8\kpc$.
It is convenient to define the dimensionless control parameters
\begin{equation*}
  R_\alpha\equiv\frac{\alpha\f h}{\etat},\quad R_\omega\equiv\frac{Gh^2}{\etat},\quad R_U\equiv\frac{U\f h}{\etat},\quad R_\kappa\equiv\frac{\kappa}{\etat},
\end{equation*}
where $G=r\,\del\Omega/\del r$.
Using equations \eqref{Krause} for $\alpha\f$ and \eqref{etat} for $\etat$, 
the dynamo number is obtained as
\begin{equation*}
\label{D}
D\equiv R_\alpha R_\omega = \frac{\alpha\f G h^3}{\etat^2} \simeq -9\frac{h^2\Omega^2}{u^2},
\end{equation*}
where the last equality applies to a flat rotation curve, $G=-\Omega$. 
Radially-dependent parameter values are given in Table~\ref{tab:models}.

%-----------------------------------------------------------------------------------------
\begin{table*}
\begin{center}
\caption{Key parameter values for the models studied.
         From left to right, 
         the radius in the disc $r$,
         the disc scale height $h$,
         the vertical turbulent diffusion time $t\diff=h^2/\etat$,
         the equipartition field strength $B\eq$,
         the amplitude of the $\alpha$ effect $\alpha\f$,
         the radial shear $G=r d\Omega/dr$,
         the amplitude of the outflow velocity $U\f$,
         and the turbulent diffusivity (of $\alpha\magn$) coefficient $\kappa$.
         This is followed by the dimensionless control parameters
         $R_\alpha\equiv\alpha\f h/\etat$,
         $R_\omega\equiv G h^2/\etat$,
         $R_U\equiv U\f h/\etat$,
         $R_\kappa\equiv \kappa/\etat$
         and $D\equiv R_\alpha R_\omega$.
         For the reader's convenience, both dimensionless and dimensional parameters are provided, e.g. $R_U$ and $U\f$.}
\label{tab:models}
\begin{tabular}{@{}ccccccccccccc@{}}
\hline
$r$        &$h$       &$t\diff$    &$B\eq$  &$\alpha\f$   &$G$           &$U\f$      &$\kappa$     &$R_\alpha$ &$R_\omega$ &$R_U$      &$R_\kappa$ &$D$     \\
$[\!\kpc]$ &$[\!\pc]$ &$\left[\!\Myr\right]$ &$[B\f]$ &$\left[\frac{\!\km}{\!\s}\right]$   
&$\left[\frac{\!\km}{\!\s\kpc}\right]$ &$\left[\frac{\!\km}{\!\s}\right]$ &$\left[\frac{\!\km\kpc}{\!\s}\right]$ &     &     &     &     &        \\
\hline                                                                                                                                        
$4$      &$381$   &$425$     &$0.82$  &$1.50$     &$-45.6$     &$0$/$1$  &$0$/$0.1$  &$1.71$     &$-19.8$    &$0$/$1.14$ &$0$/$0.3$  &$-33.9$ \\
\hline                                                                                                                                        
$8$      &$453$   &$601$     &$0.67$  &$0.68$     &$-29.1$     &$0$/$1$  &$0$/$0.1$  &$0.93$     &$-17.9$    &$0$/$1.36$ &$0$/$0.3$  &$-16.6$ \\
\hline                   
\end{tabular}
\end{center}
\end{table*}

%----------------------------------------------------------------------------------
\section{Basic equations}
\label{sec:equations}
We solve equations \eqref{dynamo} and \eqref{dalpha_mdt} in the thin-disc approximation
($h\ll r_\omega, r_\mathrm{D}, R$, hence $|B_z|\ll|B_r|, |B_\phi|$), 
where radial derivatives of $\meanv{B}$ are neglected,
and assume $\eta\turb=\const$.
In cylindrical coordinates, components of equations~\eqref{dynamo} and \eqref{dalpha_mdt} become
\begin{align}
  \label{mbr}
  \frac{\del\mbr}{\del t}=&       -\frac{\del}{\del z}\left(\alpha\mbp\right) +\eta\turb\frac{\del^2\mbr}{\del z^2} -\frac{\del}{\del z}\left(\muz\mbr\right),\\
  \label{mbp}
  \frac{\del\mbp}{\del t}=& G\mbr +\frac{\del}{\del z}\left(\alpha\mbr\right) +\eta\turb\frac{\del^2\mbp}{\del z^2} -\frac{\del}{\del z}\left(\muz\mbp\right),\\
  \label{alpha_m}
  \frac{\del\alpha\magn}{\del t}=& -\frac{2\eta\turb}{l^2 B\eq^2}
                                                \left[ \alpha\left(\mbr^2+\mbp^2\right)
                                                      -\etat\left(\frac{\del\mbr}{\del z}\mbp -\frac{\del\mbp}{\del z}\mbr\right)\right]\nonumber\\
                                   & -\frac{\del}{\del z}\left(\muz\alpha\magn\right) +\kappa\frac{\del^2\alpha\magn}{\del z^2}.
\end{align}
where we have neglected $\mbz\del\mup/\del z$ in equation~\eqref{mbp}, 
and assumed $\mbz^2\ll\mbr^2+\mbp^2$ in equation~\eqref{alpha_m}.
The solenoidality of $\meanv{B}$ implies that $\del\mbz/\del z\approx0$ in a thin disc.
Vacuum boundary conditions, $\mbr=\mbp=0$ and $\del^2\alpha\magn/\del z^2=0$ at $z=\pm h$, are used.
Under such conditions, the quadrupole mode, 
such that $\del\mbr/\del z=\del\mbp/\del z=\alpha\magn=0$ at $z=0$
emerges automatically.
The boundary condition for $\alpha\magn$ leaves unconstrained the helicity flux through the disc surface.

%------------------------------------------------------------------------
\subsection{Solutions of the disc dynamo equations}
%---------------------------------------------------------------------
\subsubsection{Dynamical non-linearity}
\label{sec:dynq}
Equations \eqref{mbr}--\eqref{alpha_m} are solved on a grid of 201 gridpoints in $-h\le z \le h$
using the 6th order finite differencing and 3rd order time-stepping schemes given in \citet{Brandenburg03}.
The seed field is taken to be $\mbr=10^{-3}B\f(1-z^2/h^2)\exp(-z^2/h^2)$, $\mbp=0$ at $t=0$, 
and $\alpha\magn=0$ initially.
Solutions are not sensitive to the value or form of the seed field, as long as it is sufficiently weak.

%------------------------------------------------------------------------
\subsubsection{Algebraic quenching}
\label{sec:algq}
In the case of algebraic quenching, 
$\alpha\magn$ no longer enters the equations explicitly and equation \eqref{alpha_m} is not solved.
Equations \eqref{mbr} and \eqref{mbp} are solved in the same manner as above but now with $\alpha$ replaced by expression \eqref{algq},
with $B^2=\mbr^2+\mbp^2$ and $a=1$.
Subsequently, equation \eqref{algq} is refined using the no-$z$ approximation to estimate $a$, 
and the simulations are repeated using this more accurate version of algebraic quenching.

%---------------------------------------------------------
\subsubsection{Marginal kinematic solutions}
\label{sec:marginal}
If the action of the dynamical quenching is to ultimately diminish the value of $\alpha$ 
without drastically modifying its spatial variation,
then it may be reasonable to simpy rescale $\alpha$ in the kinematic problem
to a marginal value corresponding to $\del/\del t=0$.
Thus, equations \eqref{mbr} and \eqref{mbp} are solved with $\alpha=\alpha\kin$, as defined in equation \eqref{alp_k}.
However, the right hand side of equation \eqref{Krause} is multiplied by a positive numerical factor $<1$, 
which is varied iteratively until the growth rate of $\meanv{B}$ reduces to zero.

%--------------------------------------------------
\subsubsection{Perturbation solutions}
\label{sec:pert}
It can be useful to have an analytic expression for $\meanv{B}$ in the kinematic regime,
and such a solution has been derived using perturbation theory for the case $\muz=0$
 \citep{Shukurov04,Sur+07,Shukurov+Sokoloff08}.
This solution is extended to $\muz\ne0$ in Appendix \ref{sec:pert_app}.
For the exponential growth rate we obtain
\begin{equation}
  \label{gammapert}
    \gamma=-\frac{\pi^2}{4} +\frac{\sqrt{-\pi D}}{2} -\frac{R_U}{2} +\frac{3\sqrt{-\pi D}R_U}{4\pi(\pi+4)}
           +\frac{R_U^2}{2\pi^2}\left(1 -\frac{\pi^2}{6}\right).
\end{equation}
As in Section \ref{sec:marginal}, the corresponding marginal solution is obtained
by using the critical values for the dynamo number $D$ and $R_\alpha=D/R_\omega$:
\begin{equation}
  \label{Brpert}
  \begin{split}
    \mbr= &C\f R_{\alpha,\mathrm{c}}\Bigg\{\cos\left(\frac{\pi z}{2}\right) +\frac{3}{4\pi^2}\left( \sqrt{-\pi D\crit} 
            -\frac{R_U}{2}\right)\cos\left(\frac{3\pi z}{2}\right)\\
           &\quad\left.+\frac{R_U}{2\pi^2}\displaystyle\sum^\infty_{n=2}\frac{(-1)^n(2n+1)}{n^2(n+1)^2}
            \cos\left[\left(n+\tfrac{1}{2}\right)\pi z\right]\right\},
  \end{split}
\end{equation}
\begin{equation}
  \label{Bppert}
  \begin{split}
    \mbp= &-\frac{2}{\pi}C\f\sqrt{-\pi D\crit}\Bigg\{ \cos\left(\frac{\pi z}{2}\right) 
               -\frac{3R_U}{8\pi^2}\cos\left(\frac{3\pi z}{2}\right)\\
              &\quad\left.+\frac{R_U}{2\pi^2}\displaystyle\sum^\infty_{n=2}\frac{(-1)^n(2n+1)}{n^2(n+1)^2}
               \cos\left[\left(n+\tfrac{1}{2}\right)\pi z\right]\right\},
  \end{split}
\end{equation}
where the subscript `c' denotes critical values,
$R_{\alpha,\mathrm{c}}=D\crit/R_\omega$,
$C\f$ is a normalization constant that controls the steady-state strength of the magnetic field,
and
\begin{equation}
  D\crit=-\frac{\pi^3}{4}\left\{ \frac{ 1 +2R_U/\pi^2 -( 2R_U^2/\pi^4)( 1 -\pi^2/6)}{ 1 +3R_U/[ 2\pi( \pi +4)]}\right\}^2.
\end{equation}
For practical purposes, it is sufficient to retain just a few terms in the infinite sums.

%-------------------------------------------------------------
\subsubsection{The no-$z$ approximation}
\label{sec:noz}
Finally, equations \eqref{mbr}--\eqref{alpha_m} can be solved in a steady state, $\del/\del t=0$,
in an approximate way as a set of algebraic equations
using the no-$z$ approximation \citep{Subramanian+Mestel93,Moss95,Phillips01,Chamandy+13a} 
to replace $z$-derivatives by simple divisions by $h$, 
e.g. $\del^2/\del z^2\sim-1/h^2$ and $\del/\del z\sim\pm 1/h$,
with the sign chosen appropriately.
This corresponds to using averages over the disc thickness.
Using the fact that solutions have the form $B\propto\Exp{\gamma t}$ in the kinematic regime,
we also solve for the exponential growth rate $\gamma$.
This leads to (see Appendix \ref{sec:noz_app} for details):
\begin{equation}
  \label{Dcrit}
  D\crit= -\frac{\pi^5}{32}\left(1 +\frac{1}{\pi^2}R_U\right)^2,
\end{equation}
\begin{equation}
  \label{gamma}
  \begin{split}
    \gamma&= \frac{\pi^2}{4}t\diff^{-1}\left(1 +\frac{1}{\pi^2}R_U\right)\left(\sqrt{\frac{D}{D\crit}} -1\right)\\
          &= \sqrt{\frac{2}{\pi}}t\diff^{-1}\left(\sqrt{-D} -\sqrt{-D\crit}\right),
  \end{split}
\end{equation}
\begin{equation}
  \label{psat}
  \tan p= -\left(\frac{2R_{\alpha,\mathrm{c}}}{\pi |R_\omega|}\right)^{1/2}
  	= \frac{1}{4}\,\frac{R_U+\pi^2}{R_\omega},
\end{equation}
\begin{equation}
  \label{Bsat}
  B^2= B\eq^2\frac{\xi(p)}{C}\left(\frac{D}{D\crit} -1\right)\left(R_U +\pi^2 R_\kappa\right),
\end{equation}
where $t\diff= h^2/\etat$ is the turbulent diffusion time-scale,
$p\equiv\arctan(\mbr/\mbp)$ is the magnetic pitch angle, 
$\xi(p)\equiv[1-3\cos^2p/(4\sqrt{2})]^{-1}$ and $C\equiv 2(h/l)^2$.
Note that $\gamma$ decreases linearly with $R_U$ in the no-$z$ approximation.

%-------------------------------------------------------------------------------------
\section{Results}
\label{sec:results}
\begin{figure}                     
  \includegraphics[width=\columnwidth,clip=true,trim= 0 10 0 0]{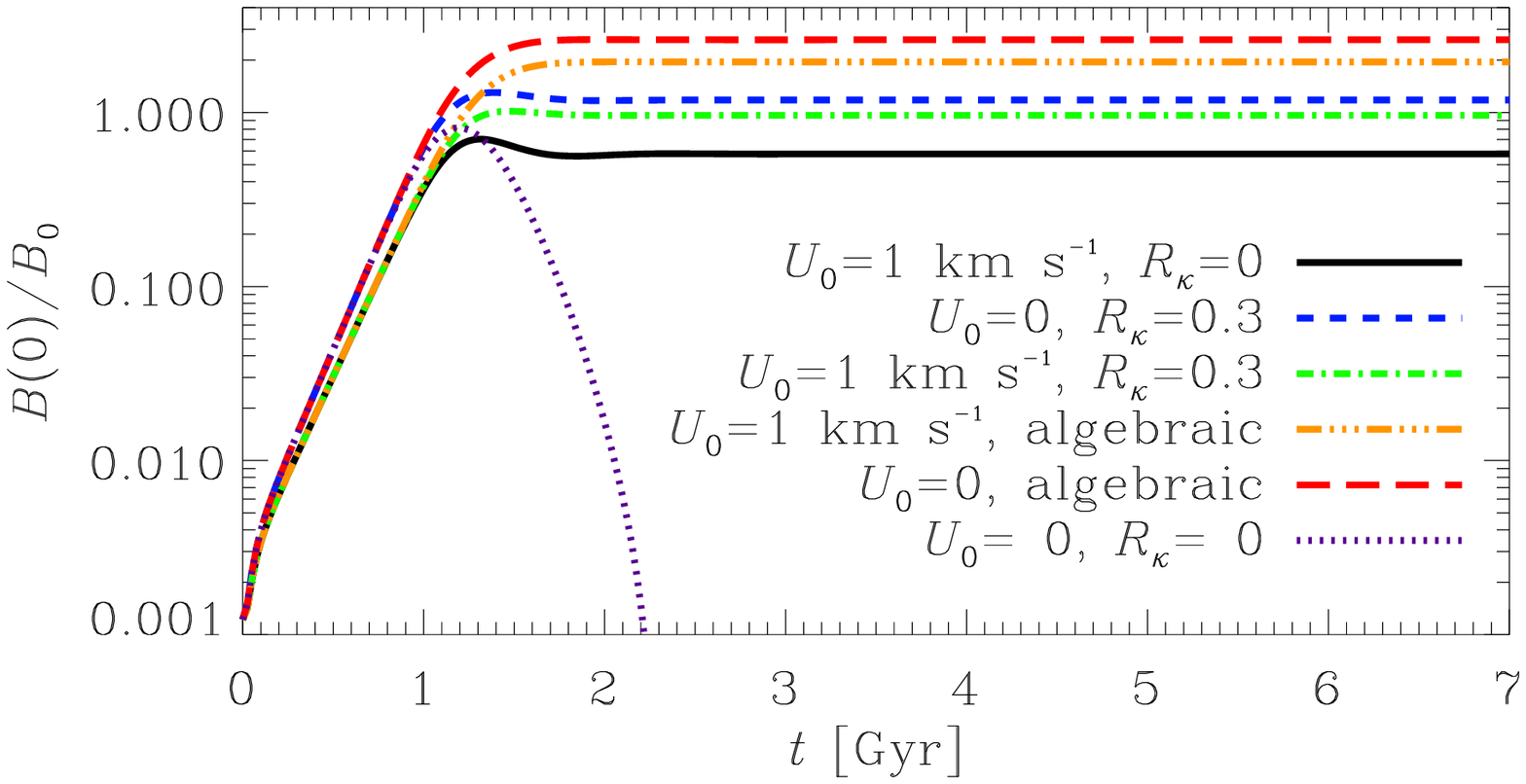}\\
  \vspace{-0.20cm}
  \includegraphics[width=\columnwidth,clip=true,trim= 0 0 0 10]{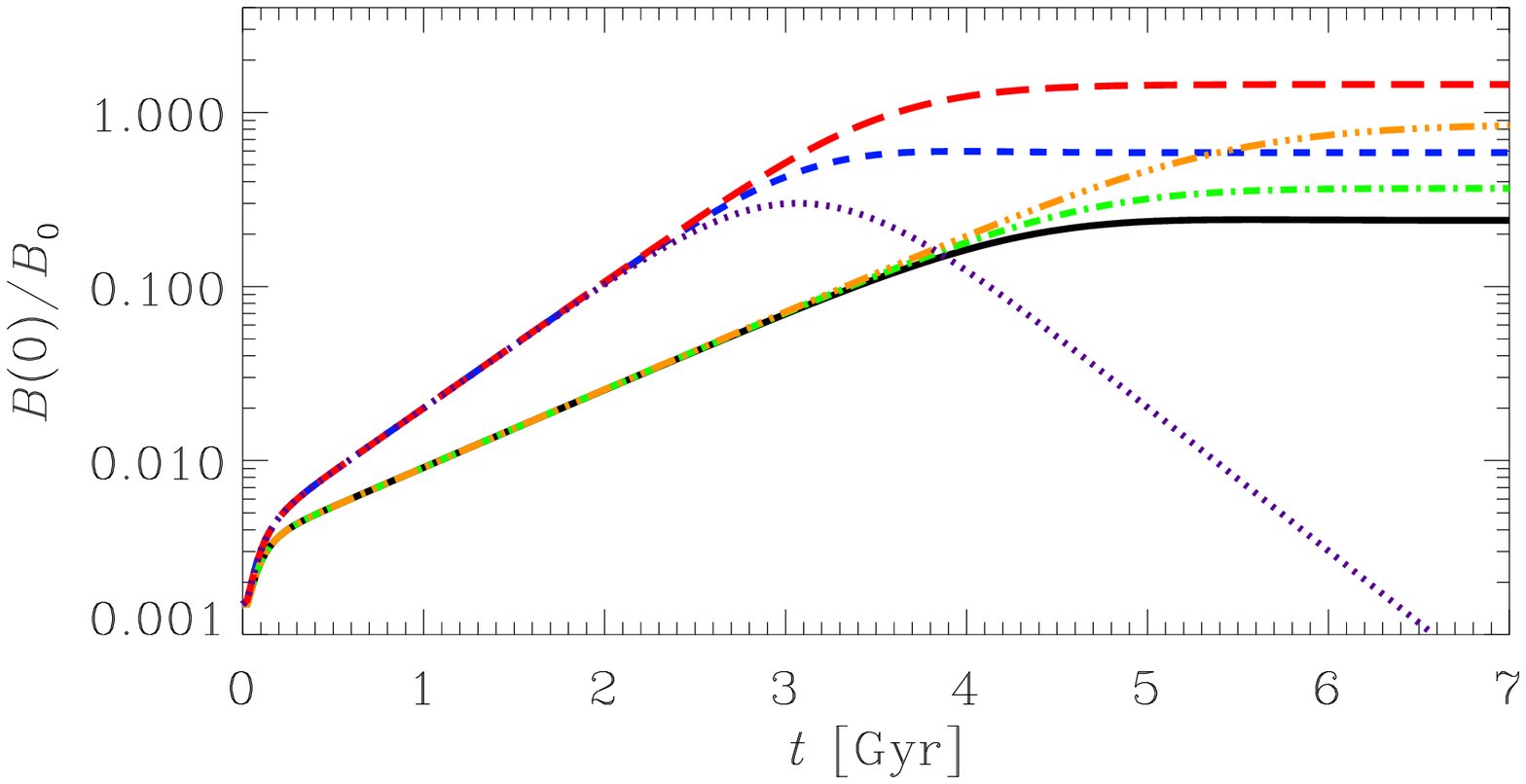}
  \caption{Evolution of the magnetic field strength at the midplane, normalized to the local equipartition field strength,
           for parameters corresponding to $r=4\kpc$ (top) and $r=8\kpc$ (bottom) in the disc model of \citet{Chamandy+13a}.
           (See online version for colour figures.)
           \label{fig:Bvstime}
          }            
\end{figure}                       

%---------------------------------------------------------------
\subsection{Growth rate, temporal evolution and saturation}
\label{sec:t_results}
The time evolution of the magnetic field strength at the galactic midplane $z=0$, 
normalized to the equipartition value $B\eq$, 
is shown in Fig.~\ref{fig:Bvstime}, for parameters corresponding to $r=4\kpc$ (top) and $r=8\kpc$ (bottom).
For solutions of the full set of equations \eqref{mbr}--\eqref{alpha_m} of Section \ref{sec:dynq},
four different regimes are depicted, with and without the advective and diffusive helicity fluxes,
$U\f=0$ or $1\kms$ and $R_\kappa=0$ or $0.3$. 
Two cases with algebraic $\alpha$ quenching (Section \ref{sec:algq}) are also illustrated, 
with $U\f=0$ and $U\f=1\kms$.

There is an initial brief phase of very rapid growth of the field at $z=0$, 
but this reflects the arbitrary choice of the seed magnetic field and is not physically relevant.
In all cases, the magnetic field then grows exponentially, until $B\sim B\eq$.
A steady state follows unless, in the dynamical quenching model, 
the flux of $\alpha\magn$ is zero ($R_\kappa=R_U=0$), 
in which case the field decays catastrophically (dotted curves).
The exponential growth rates $\gamma$ in the kinematic regime are the 
same for all solutions with a given value of $R_U$,
as would be expected.
The growth rate of the magnetic field at $r=4\kpc$ is larger than that at $r=8\kpc$ 
since the dynamo number $D$ is larger in magnitude at the smaller radius (Table~\ref{tab:models}).
The growth rate at $r=2\kpc$ (not shown here), where the eigenfunction for $\meanv{B}$ has a maximum in $r$ 
in the model of \citet{Chamandy+13a}, $\gamma=8.5\Gyr^{-1}$ for $U\f=0$,
is close (albeit slightly greater) to the global growth rate $\Gamma= 7.8\Gyr^{-1}$ 
in the axisymmetric global disc model of \citet{Chamandy+13a}.

\begin{figure} 
  \includegraphics[width=\columnwidth]{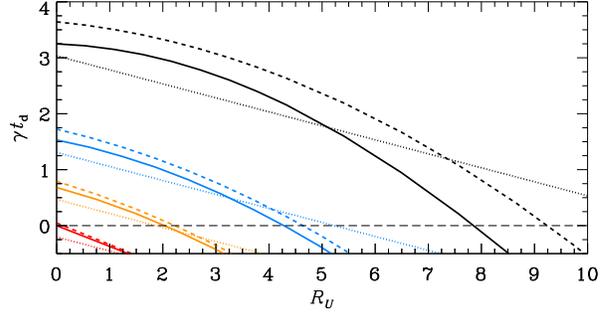}\\
  \caption{The dimensionless growth rate $\gamma$ 
           (measured in inverse diffusion time $t\diff^{-1}$)
           as a function of $R_U$ for the numerical kinematic solutions (solid),
           the perturbation solution (short-dashed),
           and the no-$z$ solution (dotted).
           Each color represents a different dynamo number/disc radius.
           From top to bottom: 
           $D=-47.5$, $t\diff=0.38\Gyr$ (black),
           $D=-22.4$, $t\diff=0.50\Gyr$ (blue),
           $D=-13.5$, $t\diff=0.73\Gyr$ (orange), and
           $D=-8.1$, $t\diff=3.67\Gyr$ (red).
           \label{fig:gamma}
          }            
\end{figure}                       

The growth rate in the numerical solutions can be compared
with that obtained from the asymptotic solutions.
Fig.~\ref{fig:gamma} shows $\gamma$, in units of the inverse diffusion time $t\diff^{-1}$,
plotted as a function of $R_U$, for various values of the dynamo number $-50<D<-8$.
Numerical solutions (solid) are well approximated 
by the perturbation solution (dashed),
and the functional form $\gamma(R_U)$ is remarkably close to that of the numerical solution.
The no-$z$ solution gives values of $\gamma$ that are somewhat less accurate 
(as might be expected) but still reasonable for growing solutions ($\gamma>0$)
unless $D$ and $R_U$ are both very large.

The steady-state field strength at the midplane,
$B(0)$, increases with $R_\kappa$ in the solutions with dynamical quenching
(compare black solid and green dash-dotted curves of Fig.~\ref{fig:Bvstime})
since larger $R_\kappa$ means larger diffusive helicity flux. 
Conversely, the saturation strength is smaller for larger $R_U$
for the values of $R_U$ considered (compare blue short dashed 
and green dash-dotted curves). 
The mean vertical velocity affects the dynamo
action in more than one way. On the one hand, the steady-state
magnetic field strength increases with $R_U$ as in equation \eqref{Bsat},
but larger $R_U$ means a larger magnitude of the
critical dynamo number \eqref{Dcrit}, hence a weaker dynamo action. As discussed
by \citet{Sur+07}, there exists a value of $R_U$ optimal for the dynamo action.
For $R_\kappa=0.3$ and $r=4\kpc$, the no-$z$ approximation has the optimal value $R_U=0.57$, 
while for $r=8\kpc$ an outflow of any intensity reduces magnetic field strength in 
the steady state (formally, the optimum value is negative, $R_U=-0.68$).
However, in the numerical solution with the dynamical quenching non-linearity 
(that does not rely on the no-$z$ approximation),
$R_U=0$ gives a higher saturation strength than any $R_U>0$ at both $r=4\kpc$ and $r=8\kpc$ for $R_\kappa=0.3$.
Similar, but less transparent dependences on $R_U$ can be noticed in the perturbation
solution of Section~\ref{sec:pert}. Similar dependence on $R_U$ also occurs under the
algebraic quenching (compare orange dash-triple-dotted and red long dashed curves),
but here the mean vertical velocity can only be damaging for the dynamo action
and the steady-state magnetic field decreases with $R_U$ monotonically.
\begin{figure*} 
  \includegraphics[width=0.50\textwidth,clip=true,trim= 10 0 85 0]{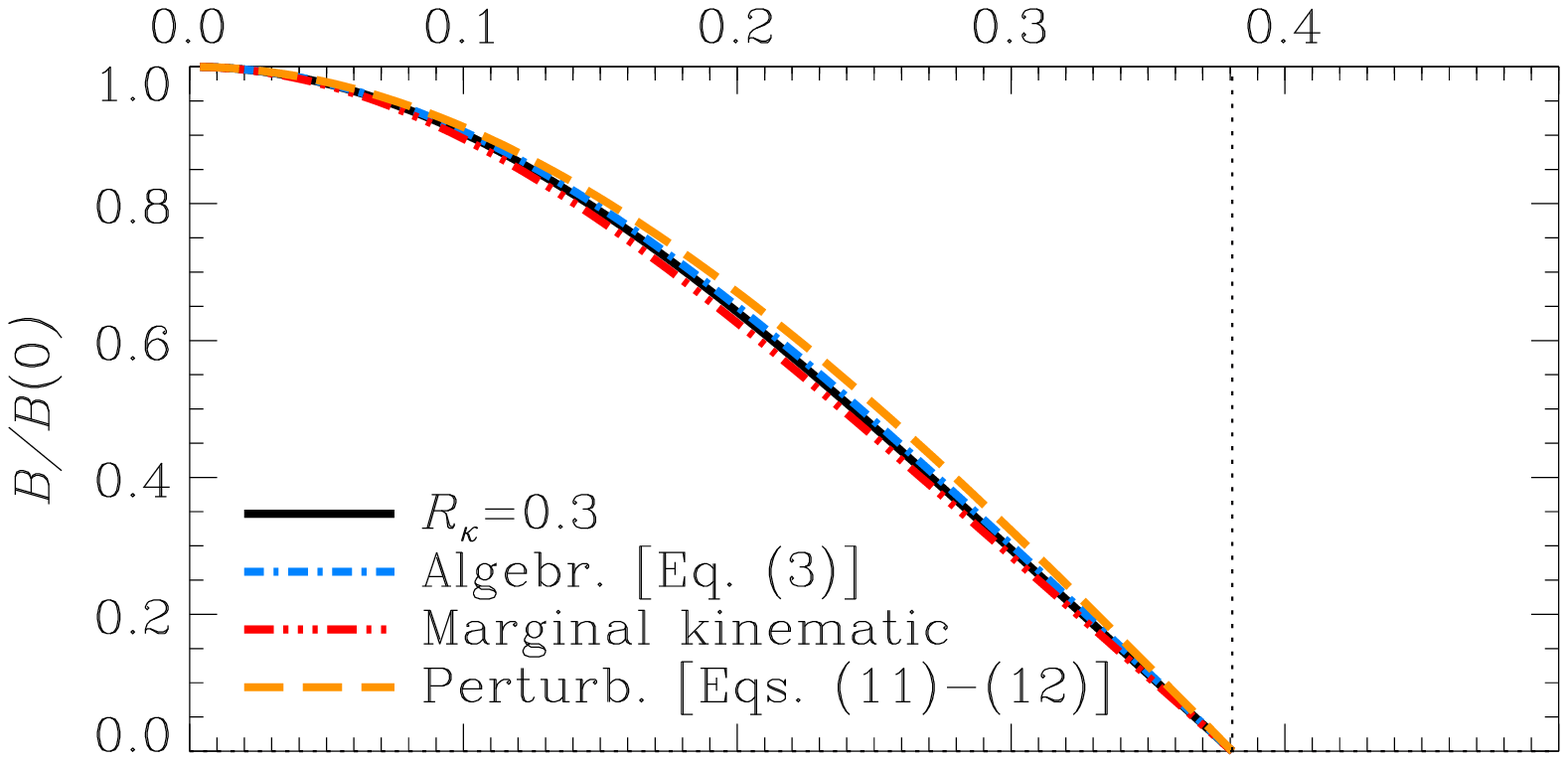}
  \hspace{-0.20cm}
  \includegraphics[width=0.50\textwidth,clip=true,trim= 85 0 10 0]{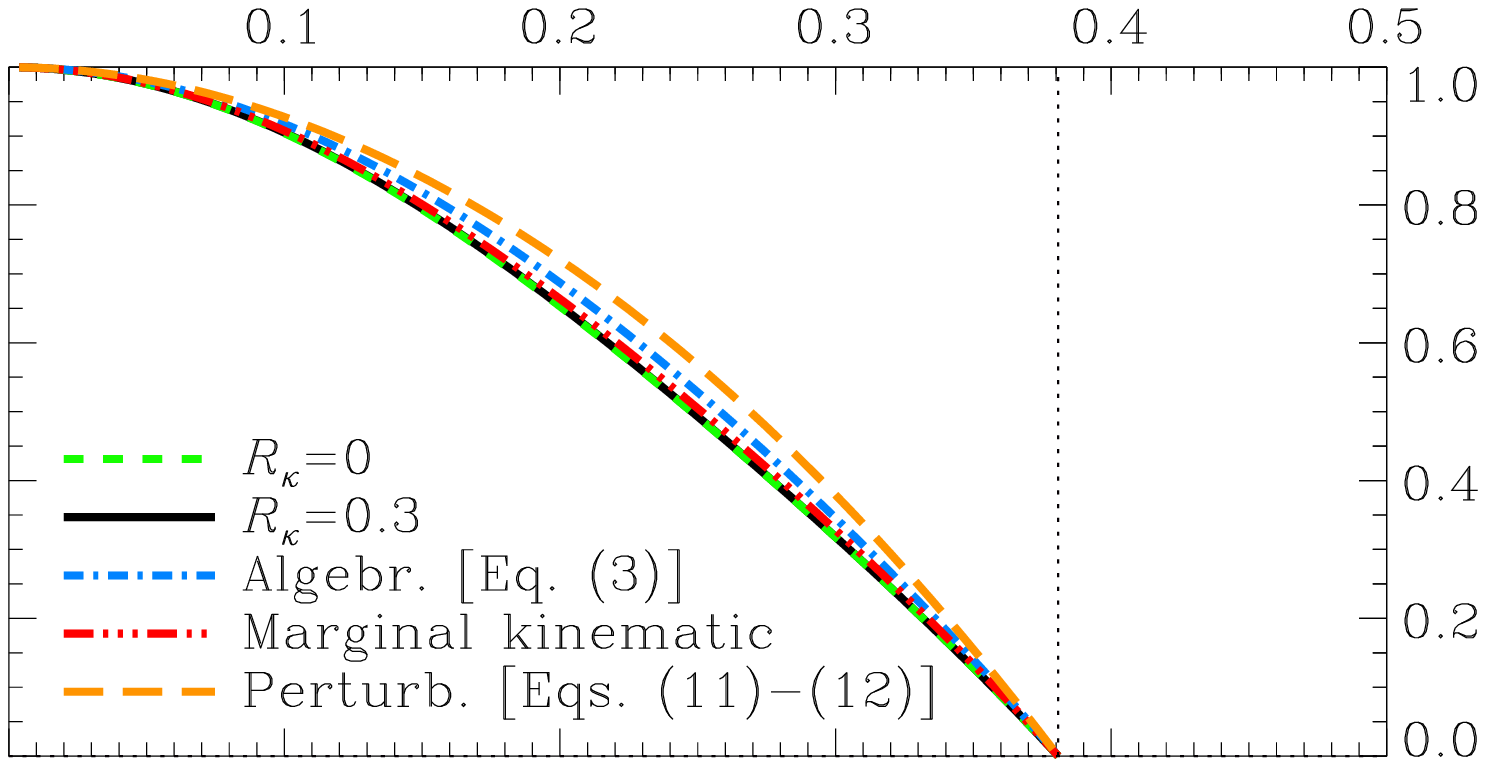}\\
  \vspace{-1.40cm}
  \includegraphics[width=0.50\textwidth,clip=true,trim= 10 0 85 0]{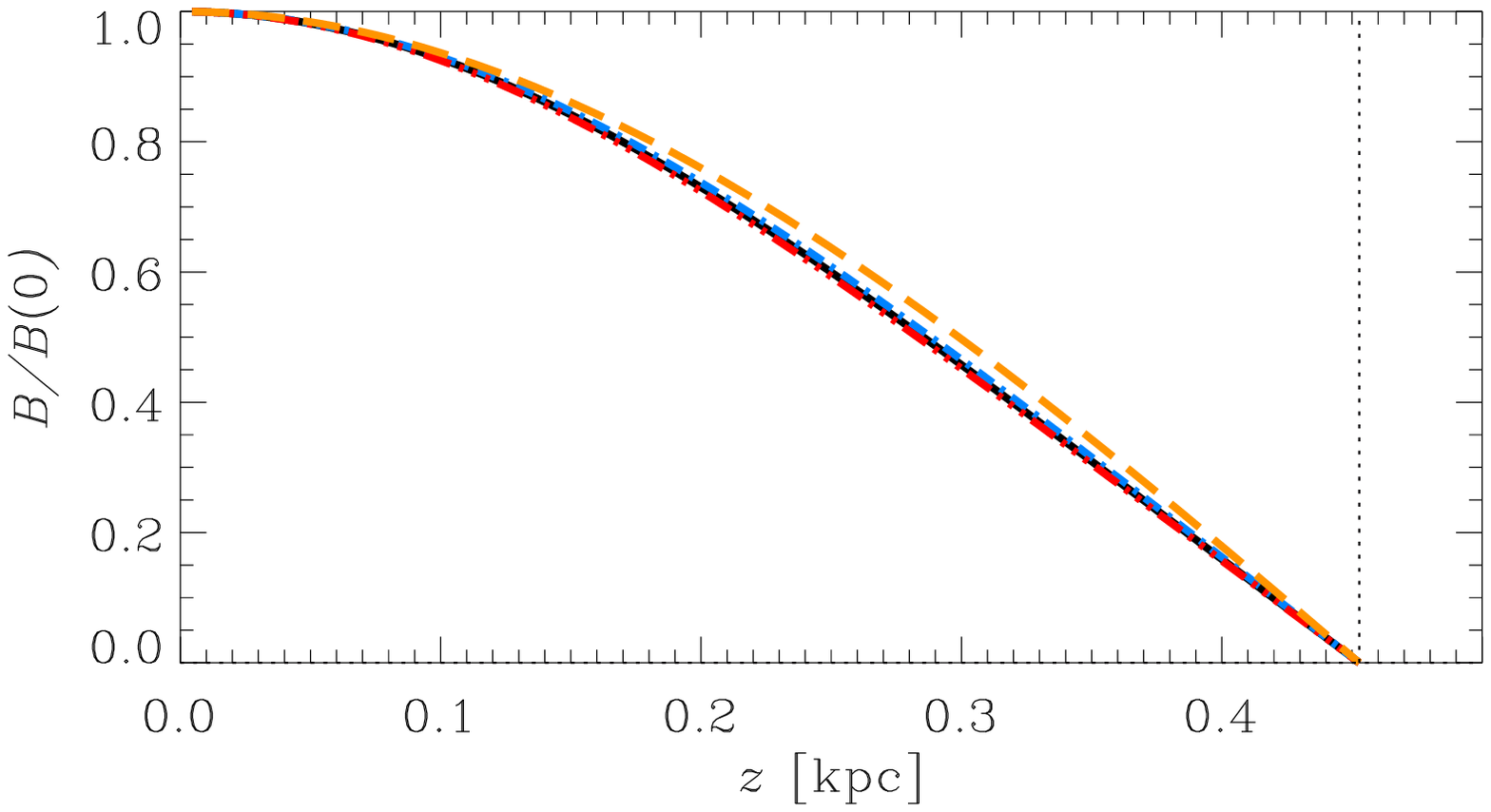}
  \hspace{-0.20cm}
  \includegraphics[width=0.50\textwidth,clip=true,trim= 85 0 10 0]{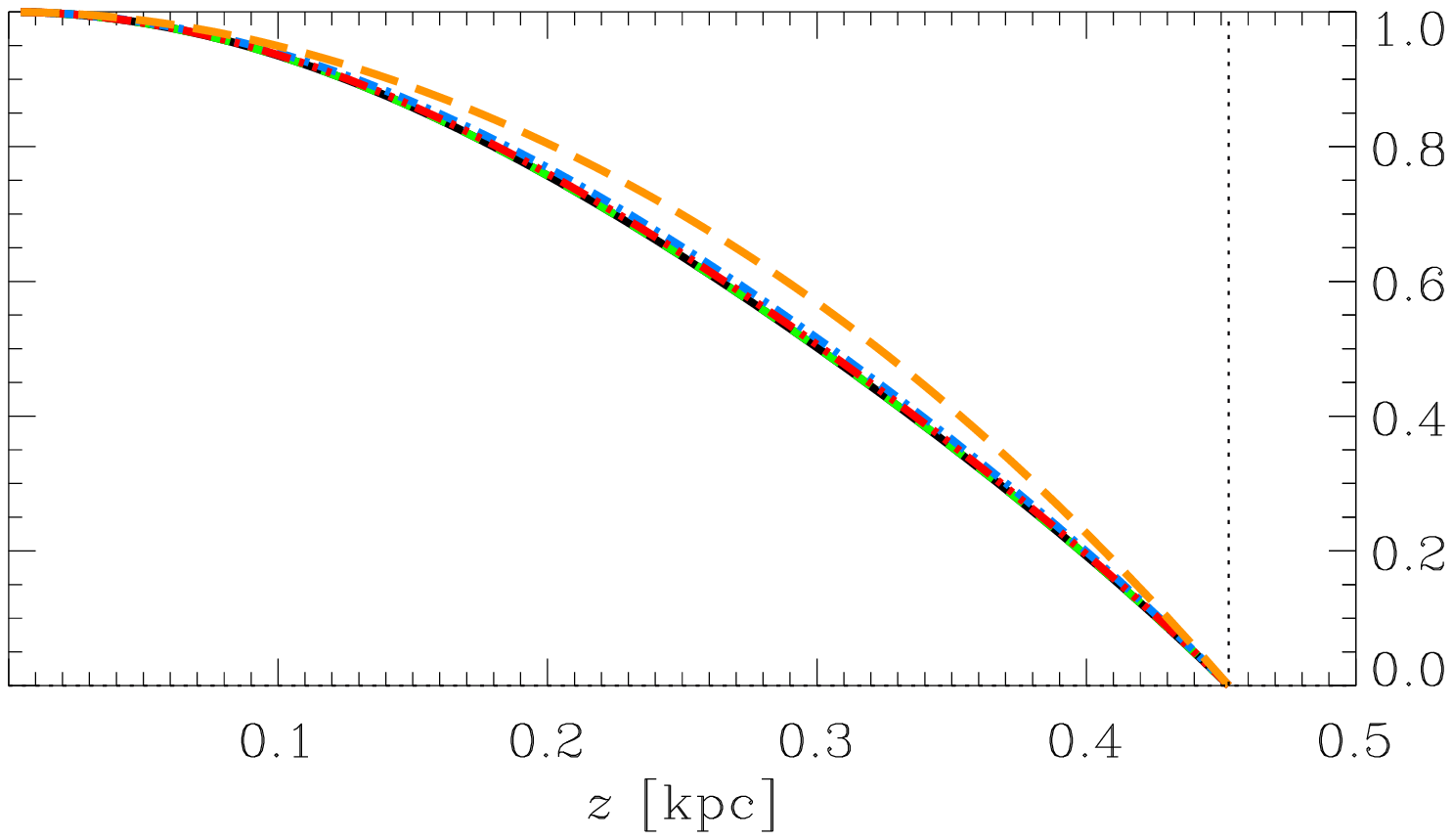}\\
  \caption{Dependence of the magnitude of the large-scale magnetic field $B$ on height $z$.
           The upper panel is for $r=4\kpc$ while the lower panel is for $r=8\kpc$.
           Solutions with $U\f=0$ are shown on the left of each panel,
           while those with $U\f=1\kms$ are shown on the right.
           The vertical dotted line shows the disk boundary at $z=h$.
           Solutions are symmetric about $z=0$.
           \label{fig:Evsz}
          }            
\end{figure*}                       

\begin{figure*}                     
  \includegraphics[width=0.50\textwidth,clip=true,trim= 10 0 85 0]{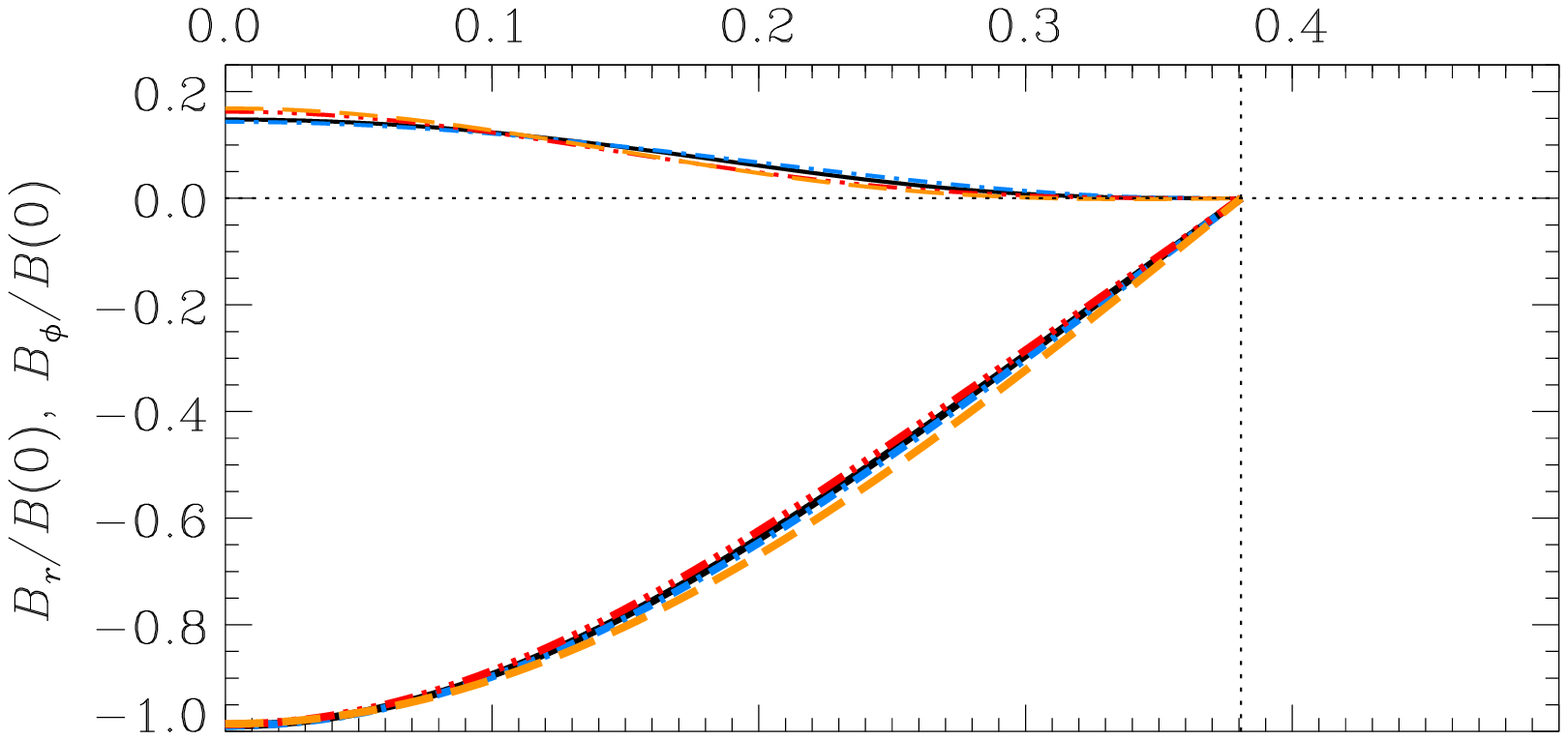}
  \hspace{-0.20cm}
  \includegraphics[width=0.50\textwidth,clip=true,trim= 85 0 10 0]{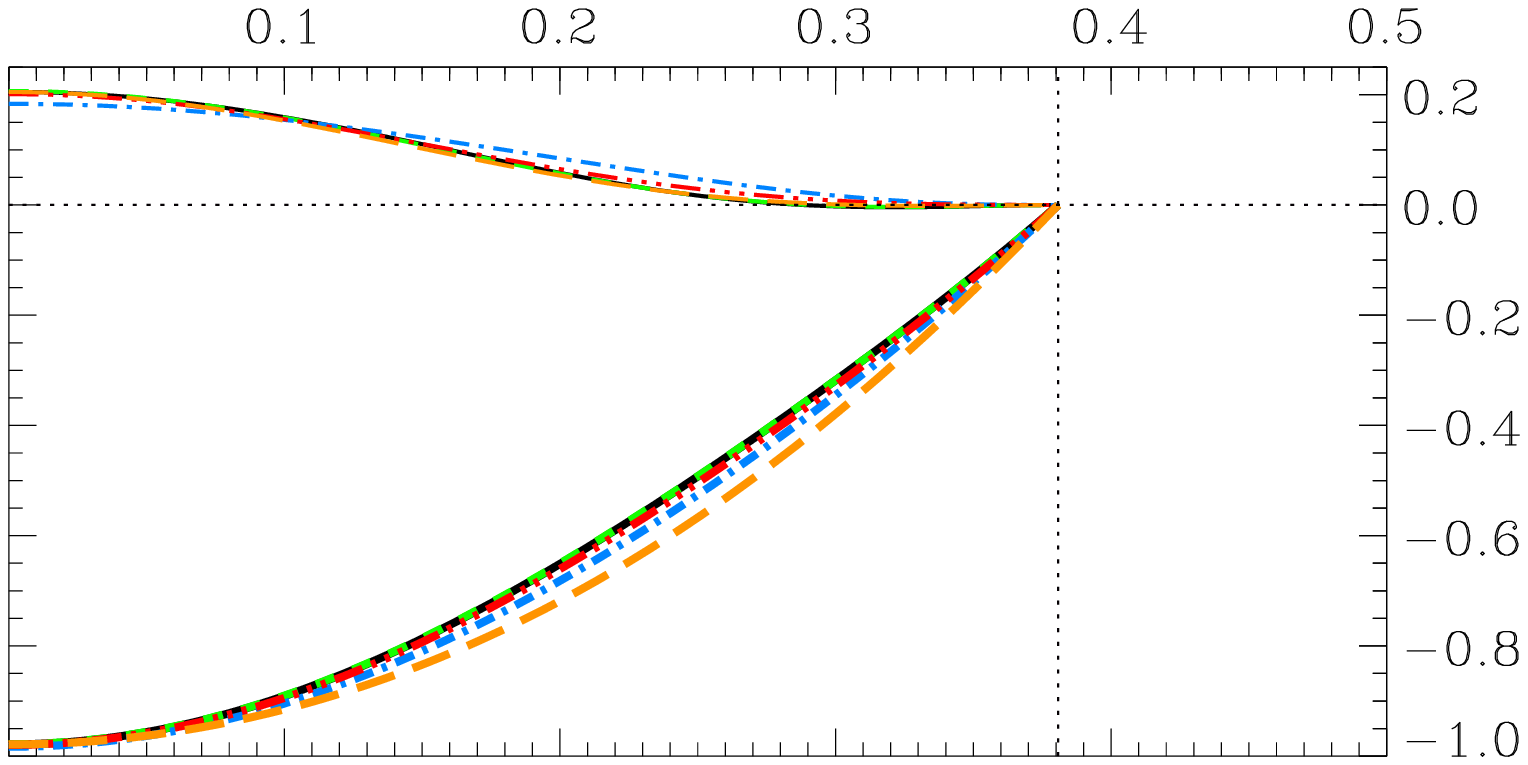}\\
  \vspace{-1.40cm}
  \includegraphics[width=0.50\textwidth,clip=true,trim= 10 0 85 0]{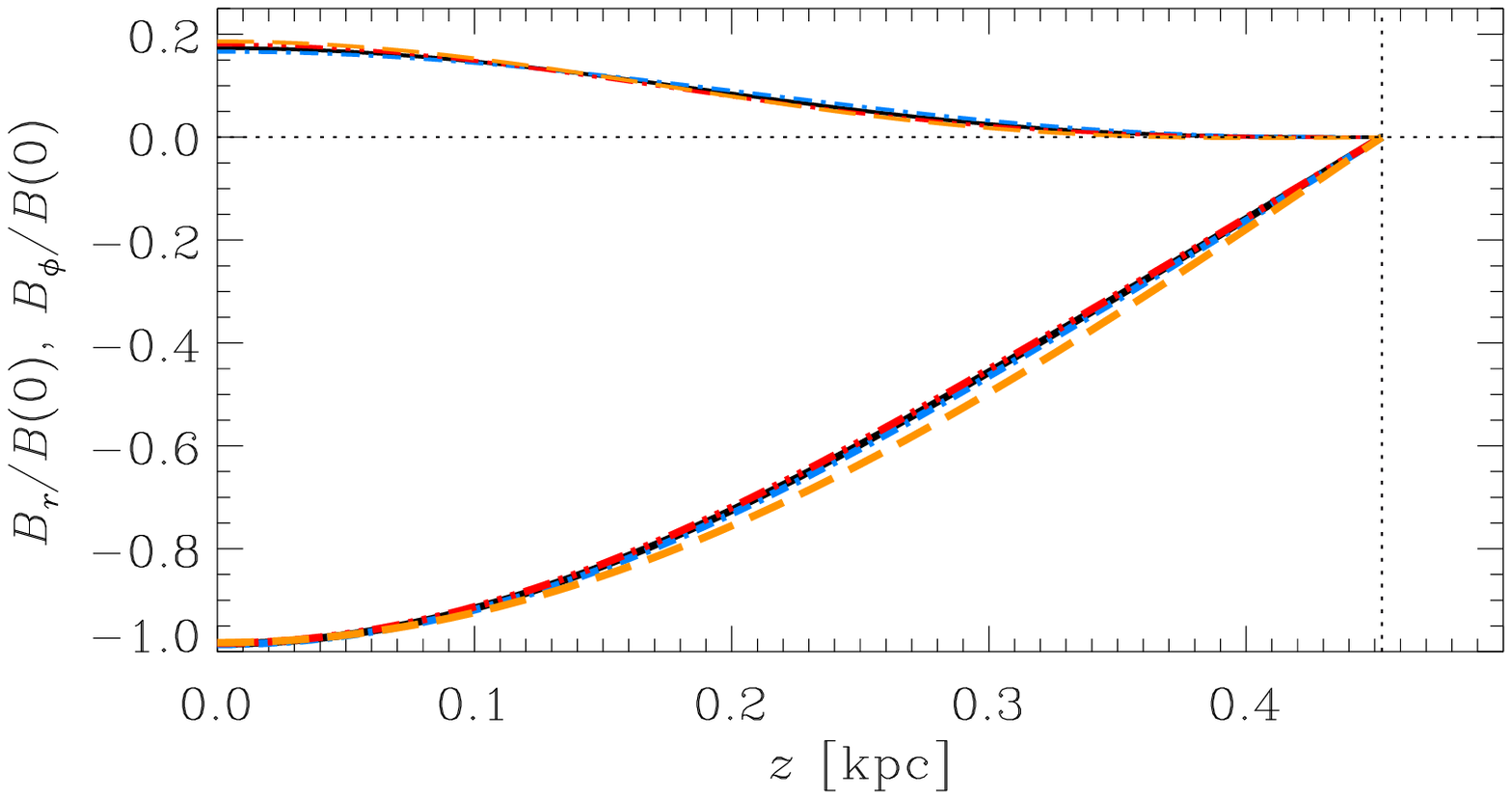}
  \hspace{-0.20cm}
  \includegraphics[width=0.50\textwidth,clip=true,trim= 85 0 10 0]{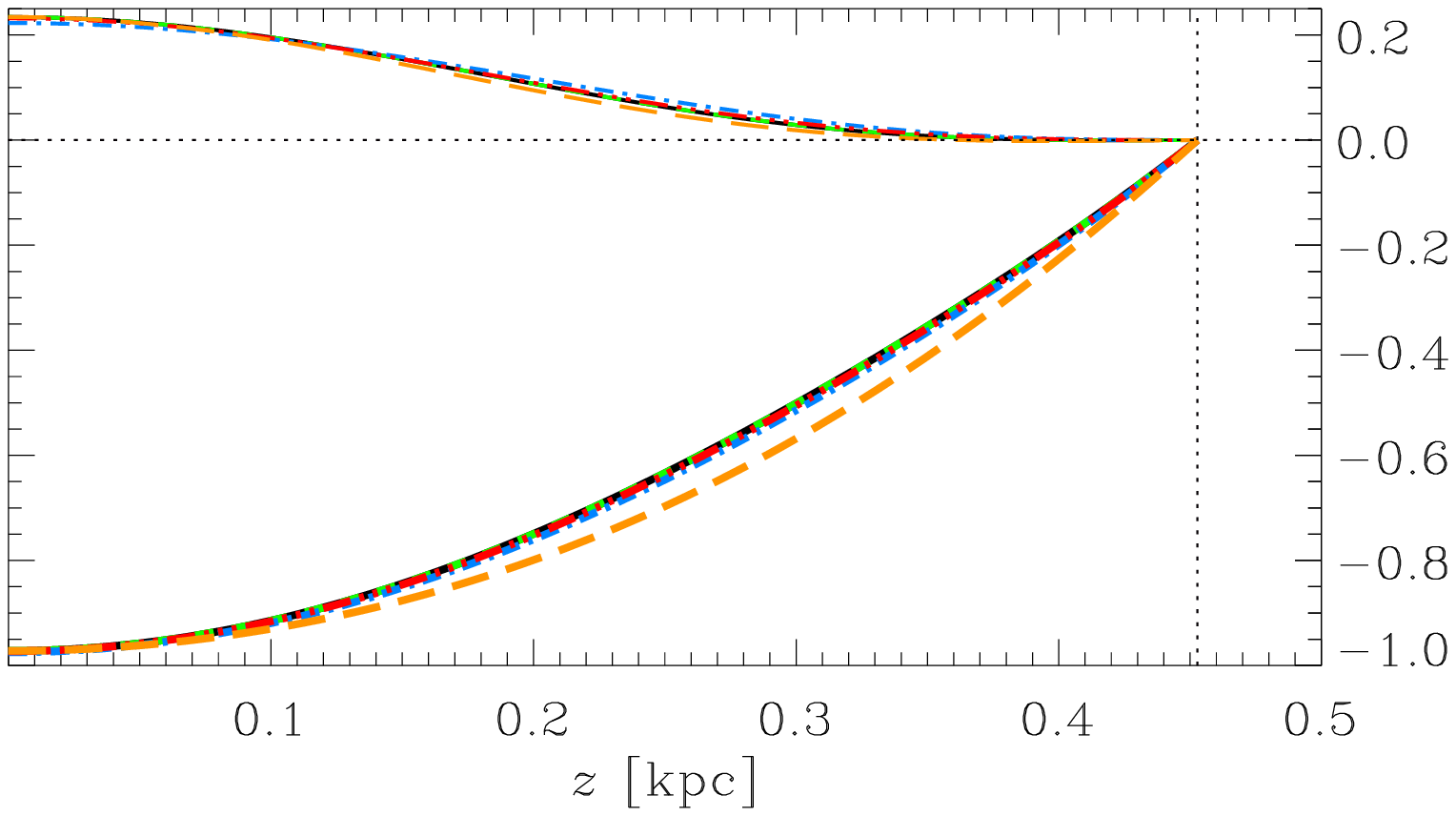}\\
  \caption{Radial (thin) and azimuthal (thick) components of $\meanv{B}$ in the saturated state, 
           normalized to the magnetic field strength at the midplane
           for parameters corresponding to $r=4\kpc$ (top) and $r=8\kpc$ (bottom)
           and to $U\f=0$ (left) and $U\f=1\kms$ (right).
           The sign of each component is arbitrary (see text), but the sign of $\mbr\mbp$ is not.
           Solutions are symmetric about $z=0$.
           (For the legend, see Fig.~\ref{fig:Evsz}.)
           \label{fig:Bvsz}
          }            
\end{figure*}                       

\begin{figure*}                     
  \includegraphics[width=0.50\textwidth,clip=true,trim= 10 0 85 0]{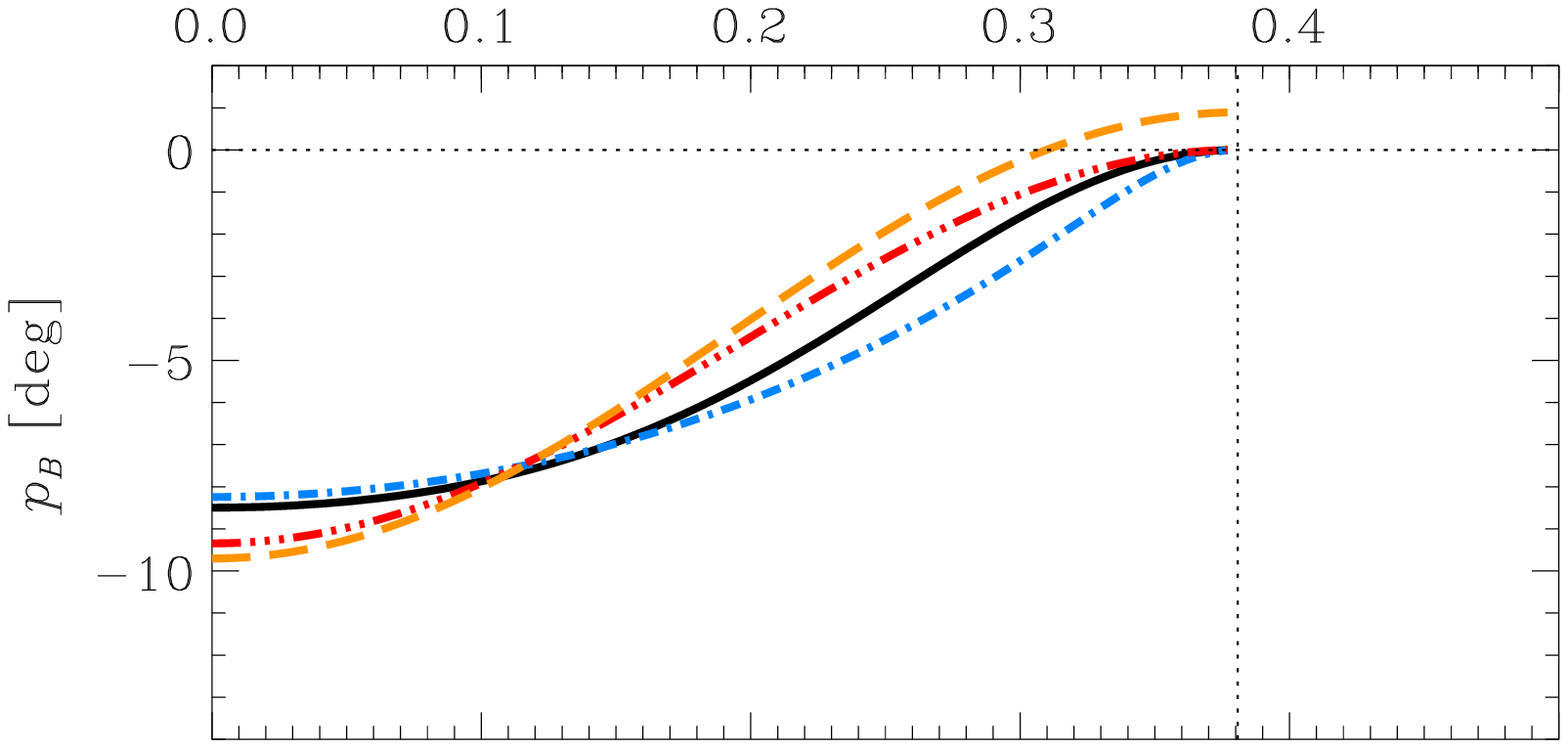}
  \hspace{-0.20cm}
  \includegraphics[width=0.50\textwidth,clip=true,trim= 85 0 10 0]{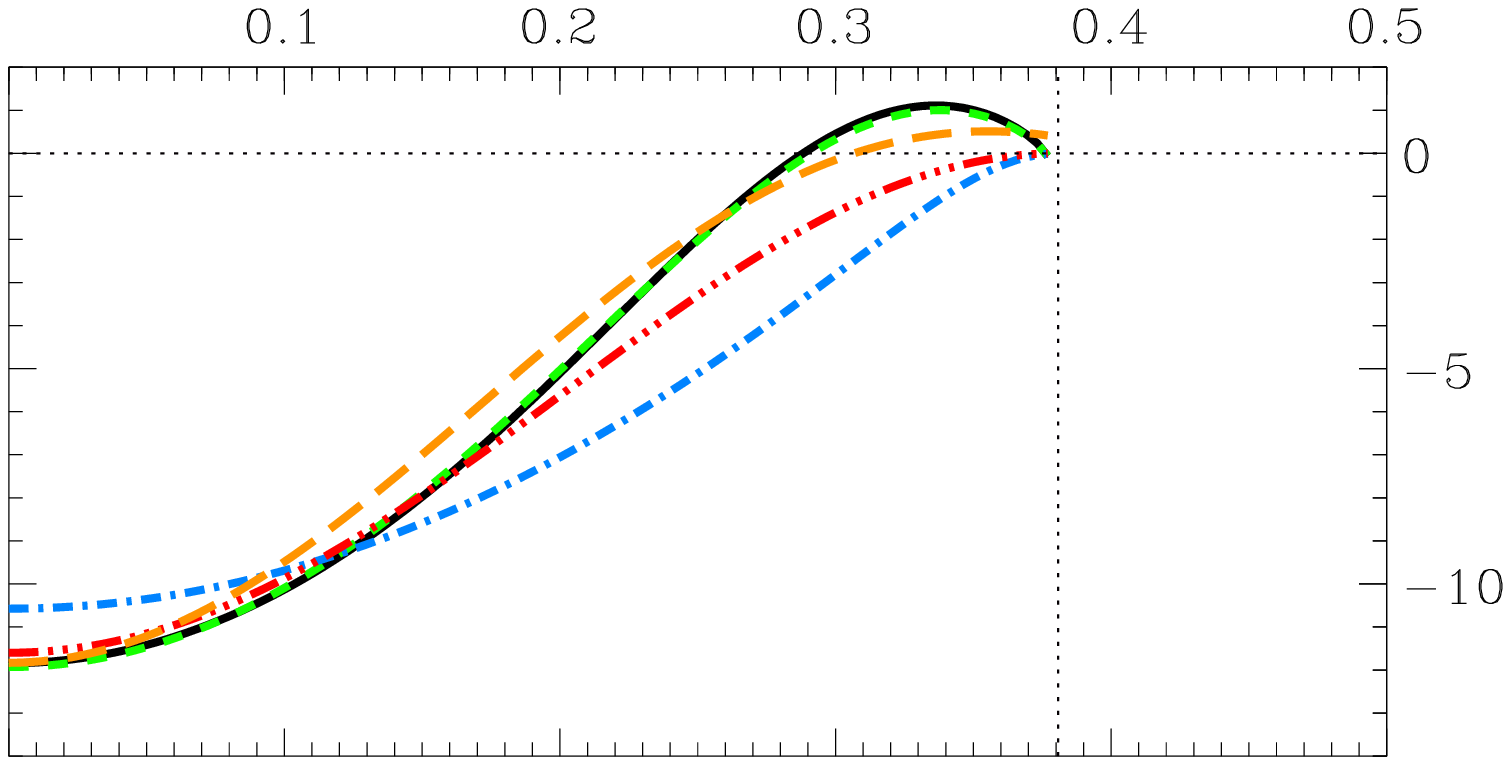}\\
  \vspace{-1.40cm}
  \includegraphics[width=0.50\textwidth,clip=true,trim= 10 0 85 0]{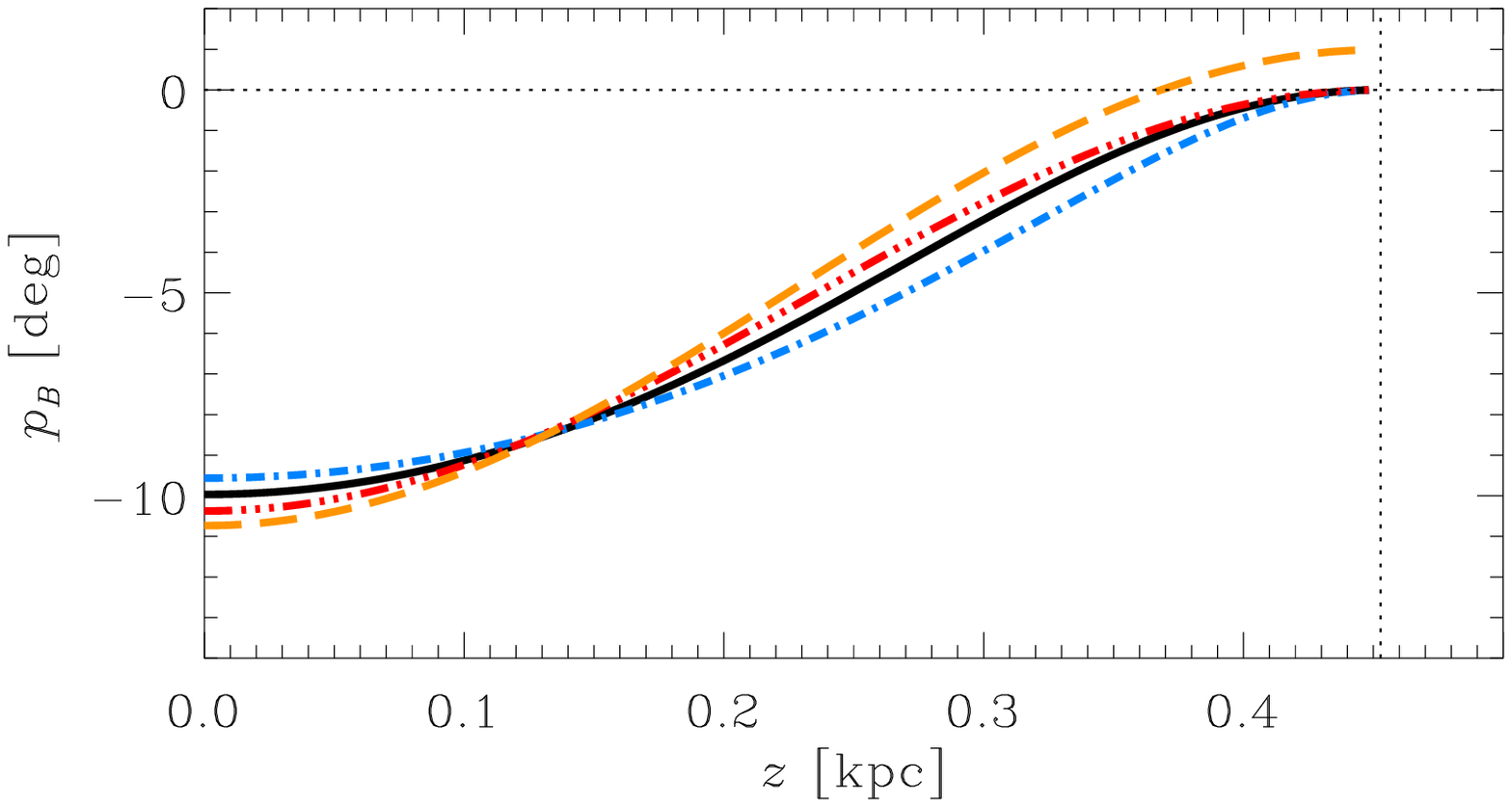}
  \hspace{-0.20cm}
  \includegraphics[width=0.50\textwidth,clip=true,trim= 85 0 10 0]{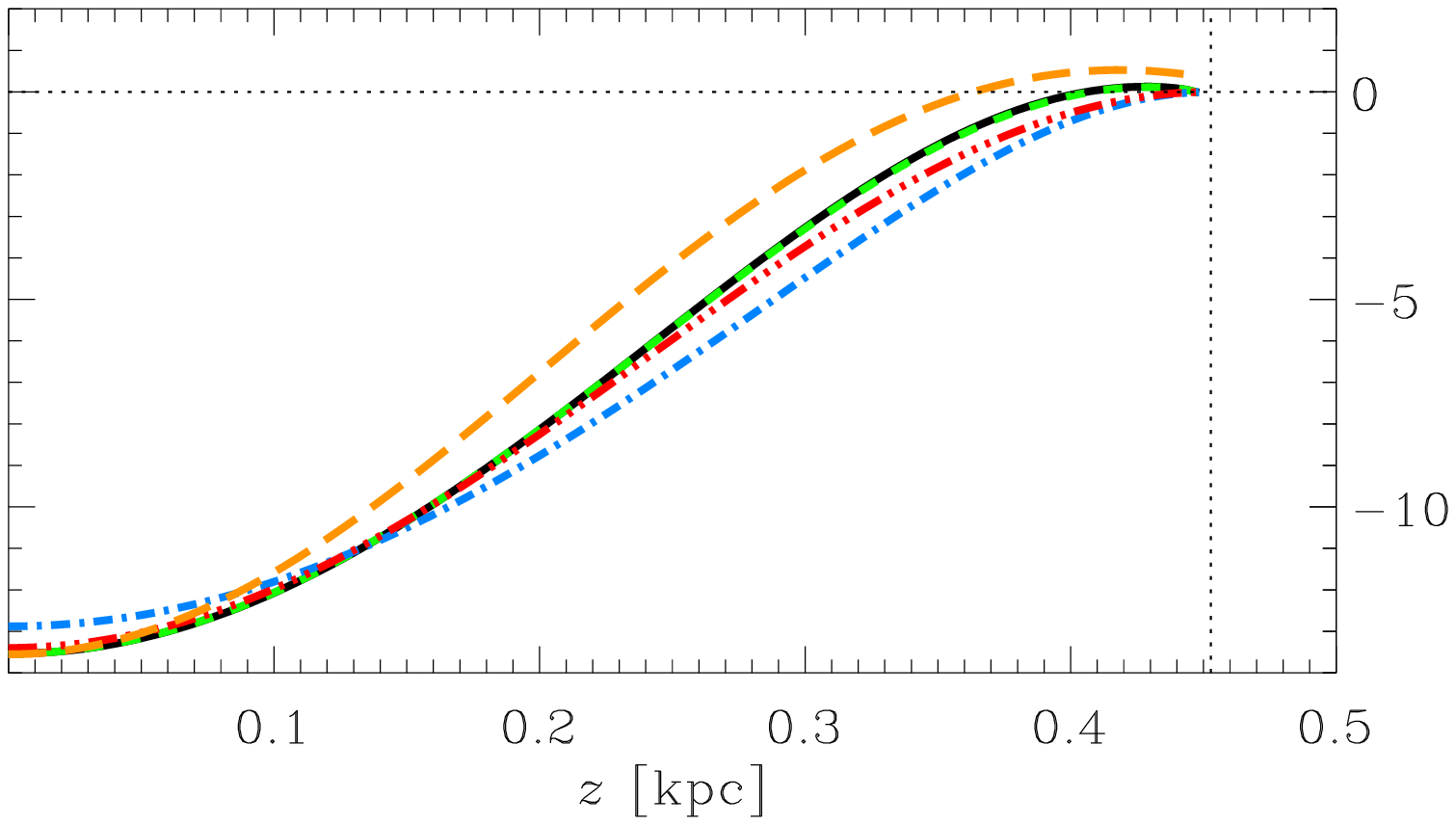}\\
  \caption{Magnetic pitch angle $p\equiv\tan^{-1}(\mbr/\mbp)$ in the saturated (steady) state, 
           as a function of the distance $z$ from the midplane,
           for parameters corresponding to $r=4\kpc$ (top) and $r=8\kpc$ (bottom)
           and to $U\f=0$ (left) and $U\f=1\kms$ (right).
           $p$ is not plotted for $z=h$, as it is undefined at the disc boundaries, 
           where the boundary conditions enforce $\mbr=\mbp=0$.
           (For the legend, see Fig.~\ref{fig:Evsz}.)
           \label{fig:pBvsz}
          }            
\end{figure*}                       

\begin{figure*}                     
  \includegraphics[width=0.50\textwidth,clip=true,trim= 10 0 85 0]{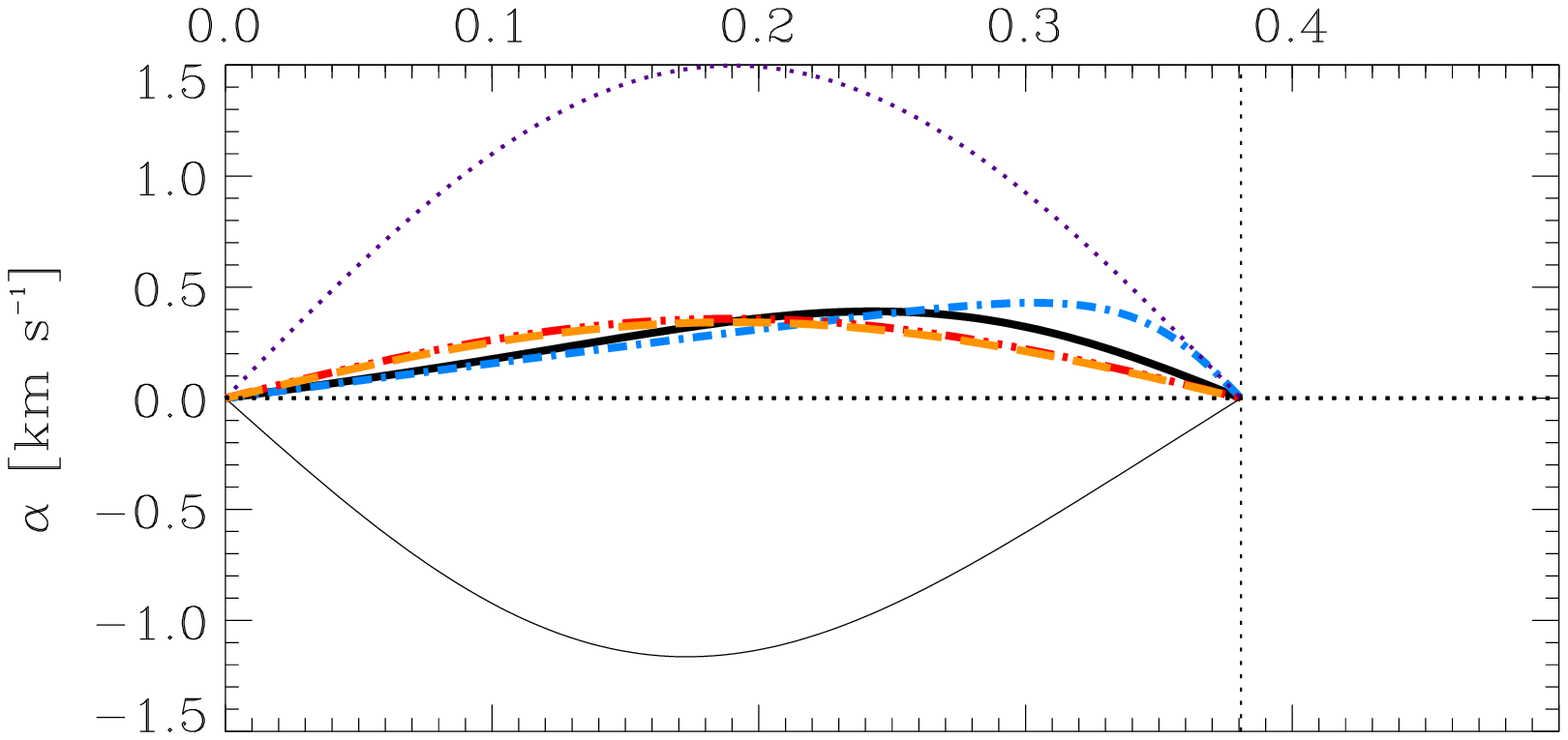}
  \hspace{-0.20cm}
  \includegraphics[width=0.50\textwidth,clip=true,trim= 85 0 10 0]{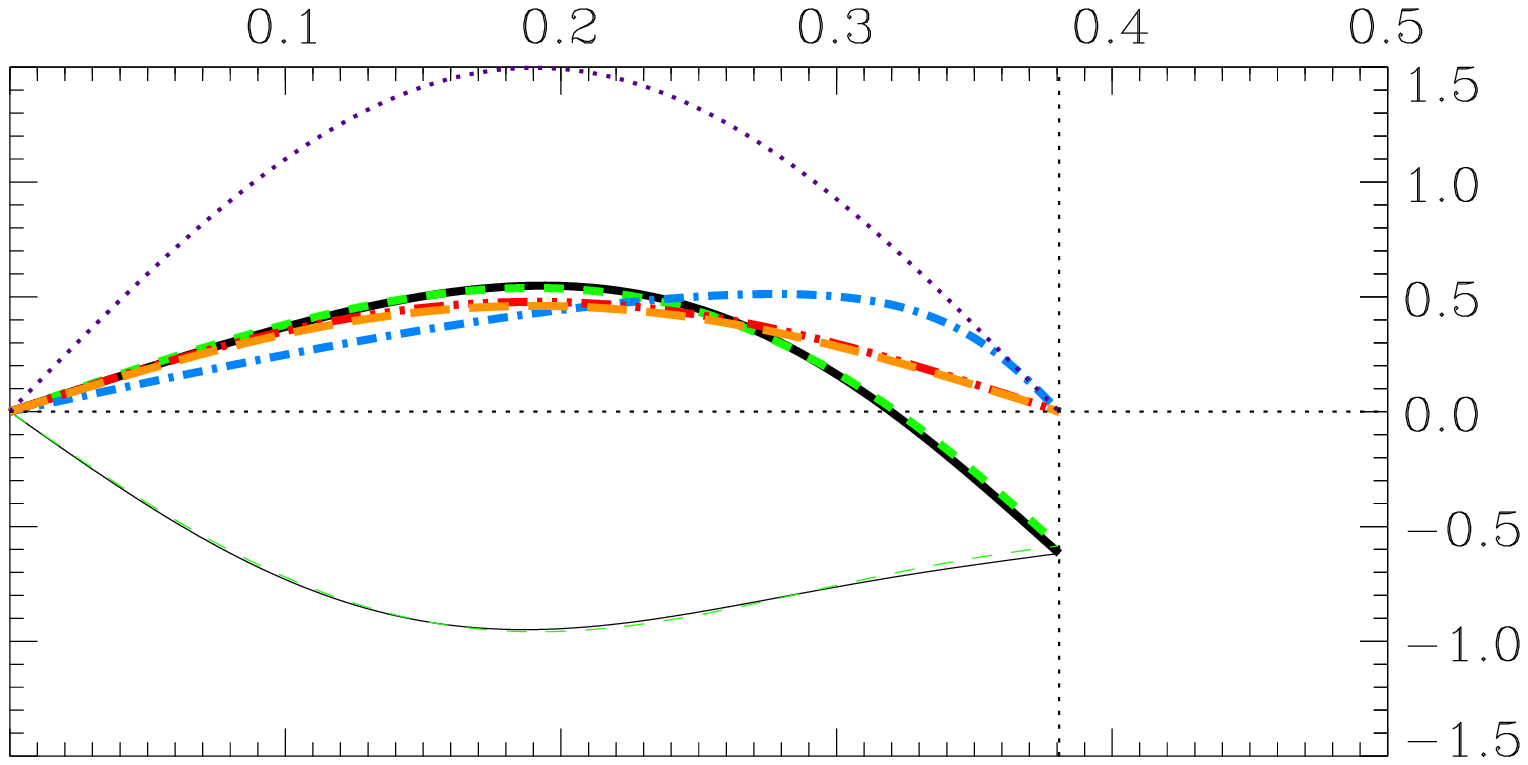}\\
  \vspace{-1.20cm}
  \includegraphics[width=0.50\textwidth,clip=true,trim= 10 -20 85 0]{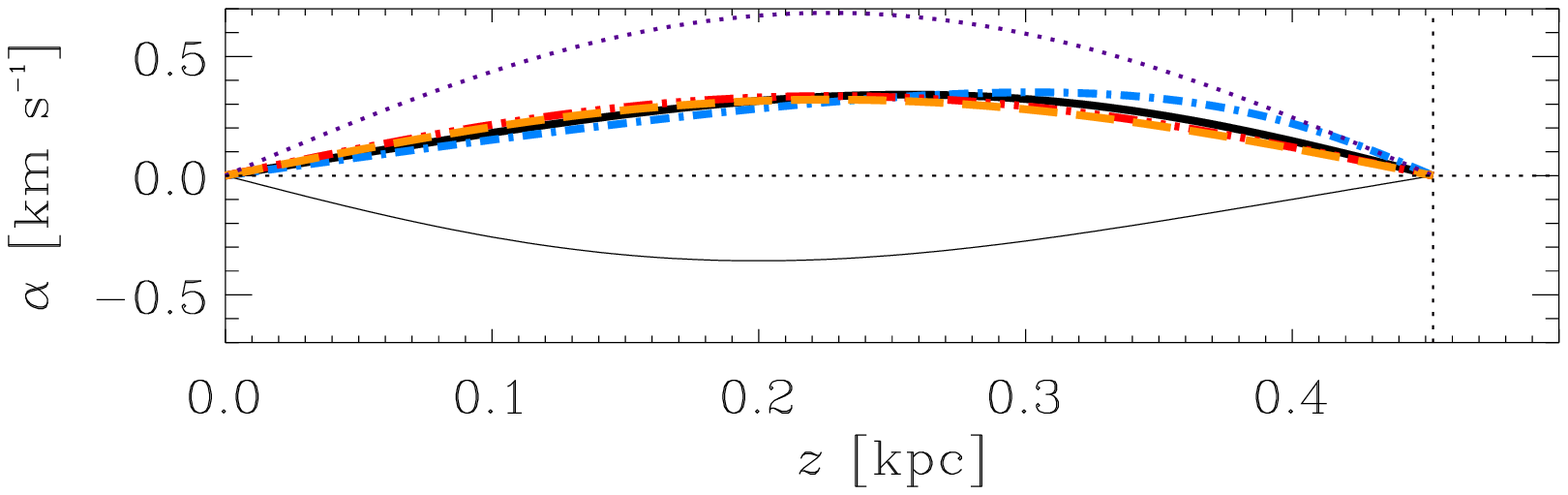}
  \hspace{-0.20cm}
  \includegraphics[width=0.50\textwidth,clip=true,trim= 85 -20 10 0]{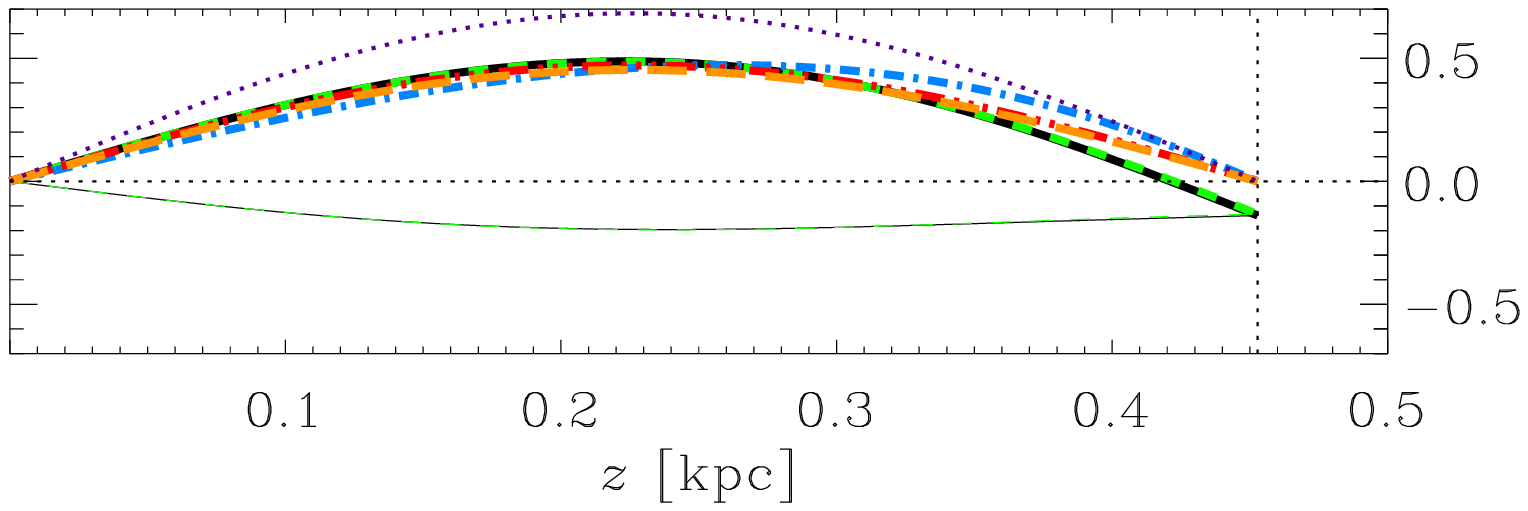}\\
  \caption{$\alpha$ in the saturated state as a function of the distance $z$ from the midplane,
           for parameters corresponding to $r=4\kpc$ (top) and $r=8\kpc$ (bottom)
           and to $U\f=0$ (left) and $U\f=1\kms$ (right).
           For the asymptotic solution, $\alpha$ has been plotted as $(\pi R_{\alpha,\mathrm{c}} \etat/2h)\sin(\pi z/h)$.
           $\alp\kin$ is shown as a thin dotted purple line for comparison.
           For solutions using the dynamical quenching non-linearity, 
           $\alpha\magn$ is shown as a thin line of the appropriate linestyle.
           All functions shown are antisymmetric about $z=0$.
           (For the legend, see Fig.~\ref{fig:Evsz}.)
           \label{fig:alpvsz}
          }            
\end{figure*}                       

With the dynamical non-linearity, 
the field undergoes mild non-linear oscillations before settling down to a steady state
(such oscillations are more evident for parameters corresponding to $r=4\kpc$, 
but are present for $r=8\kpc$ as well).
Much milder oscillations are found for the case of algebraic quenching,
for parameters corresponding to $r=4\kpc$, 
but not at $r=8\kpc$.
This oscillatory behaviour is discussed by \citet{Sur+07}
who attribute it to repeated over-suppression and recovery of the dynamo action 
by helicity fluxes.

It can be seen in both the top and bottom panels that for both $U\f=0$ and $U\f=1\kms$, 
the saturated field strength obtained using algebraic quenching is higher than that 
obtained using dynamical quenching, by a factor of about $2$--$4$.
Since the algebraic quenching is of an entirely heuristic form, this difference is
not of any physical significance. The agreement can thus be restored just by 
adjusting the factor $a$ in equation \eqref{algq}; this is done in 
Section \ref{sec:alg}.

\begin{table*}
\begin{center}
\caption{Comparison of the results of the numerical solution of Section~\ref{sec:dynq}
and the no-$z$ solution of Section~\ref{sec:noz}.
For the numerical solution, vertical averages of $p$ and $B/B\eq$,
defined in equations \eqref{p}, are given.}
\label{tab:no-z}
%\begin{tabular}{@{}ccccccccccccc@{}}
\begin{tabular}{ccccccccccccc}
\hline
                     &      &           &\phantom{.} &\multicolumn{4}{c}{Numerical solution}                                &\phantom{.} &\multicolumn{4}{c}{No-$z$ approximation}            \\
\cline{5-8}  \cline{10-13}
\phantom{$\Big($}$r$ &$R_U$ &$R_\kappa$ &\phantom{.} &$D\crit$ &$\gamma$      &$\langle p\rangle$ &$\langle B\rangle/B\eq $ &\phantom{.} &$D\crit$ &$\gamma$      &$p$        &$B/B\eq$  \\
$[\!\kpc]$             &      &           &            &         &$[\!\Gyr^{-1}]$ &$[^\circ]$         &                         &\phantom{.} &         &$[\!\Gyr^{-1}]$ &$[^\circ]$ &      \\
\hline                                                                                                                                                
$4$    &$0.0$  &$0.3$ & &$-\phantom{0}8.1$ &$5.6$  &$-6.6$ &$0.79$  & &$-\phantom{0}9.6$ &$5.1$  &$-7.1$ &$0.74$  \\
       &$1.1$  &$0.0$ & &$-10.8$           &$5.2$  &$-7.7$ &$0.40$  & &$-11.9$           &$4.5$  &$-7.9$ &$0.39$  \\
       &$1.1$  &$0.3$ & &$-10.8$           &$5.2$  &$-7.7$ &$0.66$  & &$-11.9$           &$4.5$  &$-7.9$ &$0.74$  \\
\hline              
$8$    &$0.0$  &$0.3$ & &$-\phantom{0}8.1$ &$1.7$  &$-7.2$ &$0.39$  & &$-\phantom{0}9.6$ &$1.3$  &$-7.9$ &$0.33$  \\
       &$1.4$  &$0.0$ & &$-11.5$           &$1.0$  &$-9.1$ &$0.17$  & &$-12.4$           &$0.7$  &$-8.9$ &$0.15$  \\
       &$1.4$  &$0.3$ & &$-11.5$           &$1.0$  &$-9.1$ &$0.26$  & &$-12.4$           &$0.7$  &$-8.9$ &$0.27$  \\
\hline                   
\end{tabular}
\end{center}
\end{table*}

\subsection{Magnetic field distribution across the disc}
\label{sec:z_results}
We now explore the dependence of $B$, $\mbr$, $\mbp$, $p$ and $\alpha$
on the height $z$ above the galactic midplane, in the saturated (steady) state.
In Fig.~\ref{fig:Evsz}, we plot $B$ versus $z$ for the various models, 
normalized to the value of $B$ at $z=0$ for each model,
with $r=4\kpc$ in the upper panel and $r=8\kpc$ in the lower panel.
On the left of each panel, $U\f=0$,
whereas on the right, $U\f=1\kms$.
The profiles of $\mbr$ and $\mbp$,
similarly normalized, 
are plotted in Fig.~\ref{fig:Bvsz},
while the magnetic pitch angle $p$ is shown in Fig.~\ref{fig:pBvsz}
and the $\alpha$ profile in Fig.~\ref{fig:alpvsz}.

All of the curves lie strikingly close together.
Clearly, the inclusion of a realistic outflow with $R_U\sim1$ does not drastically affect the functional form of the solution,
though some differences between solutions
with $R_U=0$ and $R_U>0$ are apparent (see below).
Furthermore, it is noteworthy that the difference between dynamical and algebraic quenching solutions is very small.
The marginal kinematic 
and perturbation solutions 
(Sections~\ref{sec:marginal} and \ref{sec:pert}, respectively)
reproduce the functional forms of $\mbr$ and $\mbp$ quite accurately
even though these are linear solutions.
The agreement is equally good in the ratio $\mean{B}_r/\mean{B}_\phi$, so all of the models 
produce 
similar magnetic pitch angles $p\sim-10^\circ$ at $z=0$ 
that reduces in magnitude with increasing distance from the midplane.

However, some small differences between the models do exist.
As shown in Fig.~\ref{fig:Evsz}, magnetic field strength $B$ 
decreases with $z$ slower as $R_U$ increases 
(compare left and right plots in each panel). 
This is more evident for $r=8\kpc$, 
where the difference between the solutions with 
$U\f=0$ and $U\f=1\kms$ is already apparent 
at $z\gtrsim0.15\kpc$.
This is a natural consequence of the advection of magnetic field
to regions with weaker field at larger $z$: such a redistribution is opposed by magnetic diffusion,
and $R_U$ is a measure of the strength of advection relative to diffusion
\citep[see also][who include a halo in their models]{Bardou+01}.

The ratio $\mbr/\mbp$ of the magnetic field components shown in Fig.~\ref{fig:Bvsz} 
is larger in magnitude near $z=0$ when $R_U\neq0$, and the ratio is larger when $R_U$ is larger.
This can be seen more clearly in Fig.~\ref{fig:pBvsz},
which shows the magnetic pitch angle $p$ as a function of $z$.
The increase of $|p|$ with $R_U$ is a direct consequence of equation~\eqref{psat}.

A notable feature of models with $R_U>0$ is that the magnetic 
pitch angle changes sign near the disc surface, so that a trailing
(with respect to the overall rotation) magnetic spiral
of the inner layers becomes a leading spiral near the disc surface.
This is a characteristic feature of any growing dynamo mode in a slab 
surrounded by vacuum \citep{Ruzmaikin+79,Ji+13},
but not of a marginal kinematic solution where $p<0$ everywhere in the slab.
The outflow extends this feature to the steady state.
The reversal of $\mbr$ responsible for this occurs deeper in the slab as the dynamo number is increased.
Thus, a leading, rather than a trailing,
magnetic spiral may exist in the disc near the disc-halo boundary (and perhaps in the halo),
provided that an outflow is present. 
However, the robustness of this feature should be checked using other types of boundary condition which may be more suitable for outflows,
or by including the galactic halo in the model.

The perturbation solution of Section~\ref{sec:pert} also has
$p>0$ near the surface for both vanishing and positive $R_U$. This is
an artefact resulting from the loss of accuracy of this solution which
is formally applicable only for $|D|\ll1$ and $R_U\ll1$. As discussed by
\citet{Ji+13}, the solution remains accurate even for $D\simeq-50$, but
this is, evidently, not the case with $R_U$. Hence, the perturbation
solution should not be used for $R_U$ of order unity if the behaviour of the solution
near the surface is important.

The change in the sign of $\alpha$ near the boundaries can be understood as resulting from the 
advective flux of $\alpha\magn$ towards the boundaries.
Since the kinetic part of the $\alpha$-coefficient is small near $z=h$,
advection of the (negative) $\alpha\magn$ from deeper layers
can change the sign of $\alpha=\alpha\kin+\alpha\magn$ near the surface.
The effect is more pronounced at smaller radius because $\alpha\magn$ 
has larger magnitude there
(see Fig.~\ref{fig:alpvsz} where $\alpha\magn$ is shown as a thin line of the appropriate linestyle and color).
It would be useful to explore how sensitive is this effect to the boundary conditions.

%------------------------------------------------------------------
\section{The algebraic non-linearity and no-\large{\MakeLowercase{$z$}} approximation refined}
\label{sec:alg}
Both the no-$z$ approximation and the algebraic form of the dynamo non-linearity are 
simple, convenient and flexible approaches, and we have demonstrated that they approximate
quite reasonably the physically motivated dynamo solutions with dynamic 
non-linearity. Then they deserve to be refined to achieve better quantitative agreement
with the results obtained under the dynamic non-linearity.

%-----------------------------------------------
\subsection{Improved no-$z$ approximation}\label{inza}
Numerical solutions of equations \eqref{mbr}--\eqref{alpha_m} are compared with
those obtained with the no-$z$ approximation in Table \ref{tab:no-z}. Since the latter
can be thought of as representing equivalent (averaged over the disc thickness) values of the
solution, we present, for the numerical solutions, the averaged magnetic field strength 
and pitch angle defined as
\begin{equation}
\label{p}
  \langle\mean{B}\rangle=\frac{1}{2h}\int_{-h}^{h} \mean{B}\,\dd z\,,
  \quad
  \tan\langle p \rangle =
  \frac{\displaystyle\int^h_{-h} \frac{\mbr}{\mbp}\, \mean{B}\,\dd z}{\displaystyle\int^h_{-h}\mean{B}\,\dd z}\,.
\end{equation}
We note that these averages may differ from those obtained from observations of
polarized intensity or Faraday rotation where the observables are the Stokes parameters
that depend on higher powers of magnetic field.

In addition to the terms used in the earlier applications of the no-$z$ approximation, 
we have additional terms representing magnetic helicity fluxes. Magnetic field components in
equations \eqref{Brpert} and \eqref{Bppert} depend on $z$ in a more complicated manner if
$R_U\neq0$. This suggests that approximating derivatives in $z$ with division by $h$ would 
be less accurate and general when $R_U\neq0$. Adjustments required
to approximate the kinematic growth rate of the mean magnetic field with a reasonable
accuracy for $-50<D<0$ are discussed in Appendix~\ref{sec:nozacc}. We suggest and use the 
following approximations:
\begin{align*}\label{refined_noz}
  \deriv{}{z}(\mean{U}_z\mean{B}_r)\simeq \frac{\mean{U}_z\mean{B}_r}{4h}\,,
  &\qquad
  \deriv{}{z}(\mean{U}_z\mean{B}_\phi)\simeq \frac{\mean{U}_z\mean{B}_\phi}{4h}\,,\nonumber\\
  \deriv{}{z}(\mean{U}_z\alpha\magn)\simeq \frac{\mean{U}_z\alpha\magn}{h}\,,
  &\qquad
  \deriv{^2\alpha\magn}{z^2}\simeq -\pi^2\frac{\alpha\magn}{h^2}\,.
\end{align*}

%---------------------------------------------------
\subsection{Algebraic $\alpha$-quenching as an approximation to the dynamic non-linearity}\label{sec:alg_quench}
The algebraic $\alpha$-quenching appears to be a reasonable approximation to
the dynamic non-linearity arising from magnetic helicity fluxes. We show in this
section that this is not a coincidence by deriving an algebraic approximation to the
dynamic non-linearity in the steady state, which contains terms responsible for the helicity
transport, and discuss conditions for the applicability of this approximation.

To explore the steady state solution, set $\del\alpha\magn/\del t=0$ in equation~\eqref{dalpha_mdt},
generalized to include the Ohmic term \citep{Sur+07}.
Assuming that the flux term can be approximated as $l^2\bmDel\cdot\Flux/(2\eta\turb) = f\alpha\magn,$ 
where $f$ is a positive numerical factor to be determined, and putting 
$\alpha\magn=\alpha-\alpha\kin$, we obtain
\begin{equation}\label{alpha}
  \alpha=\alpha\crit
        =\frac{\alpha\kin +\left(f +1/\Rm\right)^{-1}\eta\turb\bmDel\cro\meanv{B}\cdot\meanv{B}}{1+\left(f+1/\Rm\right)^{-1}B^2}.
\end{equation}
where $\alpha\crit$ is the critical value of $\alpha$ and the magnetic Reynolds number is here defined as $R\magn\equiv\etat/\eta$.
For $f=0$ (no flux of $\alpha\magn$), this reduces to equation (14) of \citet{Gruzinov+Diamond94}.\\

%----------------------------------------------------------------------------------
\begin{figure}                     
  \includegraphics[width=\columnwidth,clip=true,trim= 0 10 0 0]{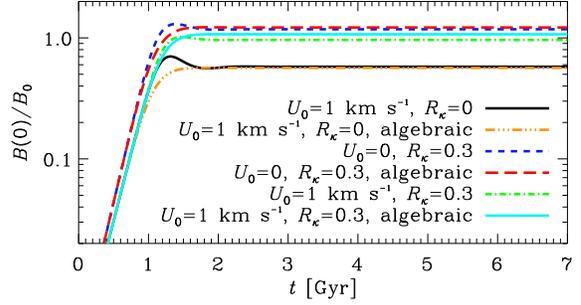}\\
  \vspace{-0.20cm}
  \includegraphics[width=\columnwidth,clip=true,trim= 0 0 0 10]{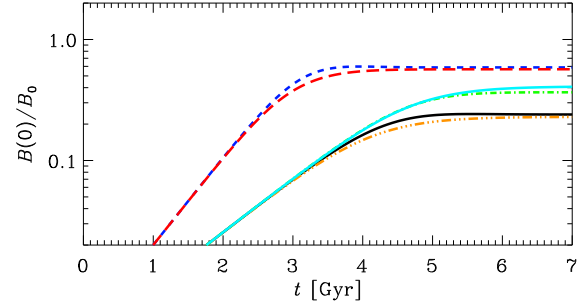}
  \caption{Time evolution of the magnetic field strength at $z=0$ for models using dynamical quenching
           and models with the same parameters but using generalized algebraic quenching with $a$ given 
           by equation \eqref{a}.
           Parameters corresponding to $r=4\kpc$ are plotted in the top panel, 
           and to $r=8\kpc$ in the bottom panel.
           \label{fig:Bvstime_gen}
          }            
\end{figure}                       

%----------------------------------------------------------------------------------

The algebraic form \eqref{algq}
follows from equation \eqref{alpha} if 
\begin{equation}
  \label{jb}
  \eta\turb\bmDel\cro\meanv{B}\cdot\meanv{B}= K\alpha\crit B^2,
\end{equation}
with $K$ a factor to be determined.
Magnetic helicity balance constrains $K$ to be positive,
since $\alpha\crit$ has opposite sign to $\alpha\magn$, 
and large- and small-scale current helicities must have opposite sign to each other.

Equation \eqref{jb} is satisfied reasonably well for typical galactic parameters.
Combining equations \eqref{jdotb} and \eqref{psat} of the no-$z$ solution, 
recalling that $D\crit=h\alpha\crit R_\omega/\etat$,
and assuming $\mbr^2/\mbp^2\ll1$, typical of $\alpha\omega$-dynamos, we obtain 
\begin{equation}
\label{K}
  K=\frac{3}{4\sqrt{2}}\approx 0.53.
\end{equation}

Furthermore, substituting $\alpha\kin/\alpha\crit$ for $D/D\crit$ in equation \eqref{Bsat},
and solving for $\alpha\crit$ (assuming $\mbr^2/\mbp^2\ll1$),
we obtain the form \eqref{algq} with
\begin{equation}\label{a}
  a=\frac{C}{\xi\f(R_U +\pi^2 R_\kappa)}\,,
\end{equation}
where $\xi\f=[1-3/(4\sqrt{2})]^{-1}\approx2.1$ and $C\equiv2(h/l)^2$,
and the numerical factors in the no-$z$ approximation estimated in Section~\ref{inza}
have been used in the helicity flux terms.
This result can also be obtained from equations \eqref{alpha}, \eqref{jb}, and \eqref{K},
with $f=(R_U +\pi^2 R_\kappa)/C$.
Specific values of $a$ at $r=4\kpc$ and $r=8\kpc$ in the galaxy model adopted 
range from $3$--$14$ for the combinations of parameter values listed in Table~\ref{tab:no-z}.
Note that this is consistent with the saturation value of $B$ being a factor of about $2$--$4$ too large 
(compared to that obtained with the dynamical non-linearity)
when using algebraic quenching with $a=1$,
as noted in Section~\ref{sec:t_results}.
Also note that these values for $a$ are somewhat sensitive to the ratio $h/l$ appearing in equation \eqref{a}.
\citet{Gressel+13a} find even larger values of $a$ from direct numerical simulations.

Equation~\eqref{a} appears as a natural generalization of the standard $\alpha$-quenching, 
as it makes the saturation magnetic field dependent on the magnetic helicity fluxes. 
Figure~\ref{fig:Bvstime_gen} compares the results obtained numerically with \eqref{algq} 
and \eqref{a} with those under dynamical quenching.
Clearly, the agreement is much better than for $a=1$ (see Fig.~\ref{fig:Bvstime}).
The $z$-distributions of the resulting magnetic field components  are identical to 
those under the standard algebraic quenching with $a=1$.

%------------------------------------------------------------------------------------
\section{Examples of application}
\label{sec:examples}
To illustrate the use of the toolbox, we apply it to estimate
the magnetic pitch angle and to explore the nature of the magnetic arms.
The analysis below is mainly based on the analytical expressions of 
the no-$z$ approximation, but results have been checked with numerical solutions.

%--------------------------------------------------------------------------
\subsection{Magnetic pitch angle}
In the kinematic stage, the pitch angle of the mean magnetic field, 
$p=\arctan(\mean{B}_r/\mean{B}_\phi)\simeq-\arctan(R_\alpha/|R_\omega|)^{1/2}$,
agrees reasonably well with observations, even if its magnitude remains
somewhat smaller than desired \citep{Ruzmaikin+88}. Some galaxies with exceptionally
open magnetic spirals with  $p\simeq-40^\circ$, such as M33 \citep{Tabatabaei+08,Chyzy+Buta08},
do not seem to be consistent with this estimate.

Non-linear dynamo effects can only reduce the magnitude of the pitch angle
since, effectively, they lead to the reduction of $R_\alpha$ to the smaller $R_{\alpha,\mathrm{c}}$
\citep[see also][]{Elstner05}.
Indeed, for the fairly typical galactic parameters chosen in the present work, 
we obtain $p\sim-10^\circ$ in the non-linear solutions,
whereas average observed values are closer to $-20^\circ$
(\citealt{Fletcher10}; see also the discussion in \citealt{Chamandy+13d}).

According to equation \eqref{psat},
$|p|$ increases with $R_U$, but increasing $R_U$ also causes $\gamma$ and $B^2$ to decrease, 
according to equations \eqref{gamma} and \eqref{Bsat},
and there is a maximum value of $R_U$ above which the dynamo action is suppressed.
From equation \eqref{Dcrit}, the dynamo remains active for 
$R_U+\pi^2<4\sqrt{(2/\pi)\,R_\alpha |R_\omega|}$, which leads to a lower limit on
the magnetic pitch angle:
\[
  \tan p >-\sqrt{\frac{2R_\alpha}{\pi|R_\omega|}}\,,
\]
so that the outflow can hardly enhance the magnitude of $p$ significantly.
The right-hand side of the equation is simply the estimate one obtains
for the kinematic regime;
for the dynamical non-linearity that is assumed, $R_{\alpha,\mathrm{c}}<R_\alpha$,
so $|p|$ always decreases from its value in the kinematic regime.

It appears that magnetic pitch angle of the dynamo-generated magnetic field is further
modified by additional effects, some of which are rather obvious. Most importantly,
magnetic field compression in the gaseous spiral arms efficiently aligns magnetic field
with the spiral arms. If the ratio of the gas densities in the arm and outside it is $\epsilon$,
with $\epsilon>1$ and $\epsilon=4$ for a strong adiabatic shock, the angle $\theta$ between magnetic field
and the arm axis is reduced, under one-dimensional compression, as 
$\tan \theta_2=\epsilon^{-1}\tan \theta_1$ from the interarm value $\theta_1$ to that within the arm $\theta_2$.
For $\theta_1=30^\circ$ and $\epsilon=4$, magnetic field is diverted by $\theta_1-\theta_2\approx22^\circ$
towards the arm axis. Since such observables as the Faraday rotation measure and polarized
synchrotron intensity are dominated by the denser and stronger magnetized interior of the spiral
arms, the compressional alignment can significantly affect the magnetic pitch angle observed.

Additional dynamo effects can also make the magnetic spirals more open. For example, 
the contribution of magnetic buoyancy to the mean-field dynamo action 
can produce $p=-(20^\circ$--$30^\circ)$ \citep{Moss+99}.
Another, less obvious effect can be due to a radial inflow of interstellar gas
(at a speed of order $U_r=1\kms$ at the Solar radius in the Milky Way, and expected 
to be stronger in galaxies with more open spiral patterns), driven
by the outward angular momentum transfer by the spiral pattern, turbulence and magnetic fields.
The inflow can increase the magnitude of $p$ by at least 5--10\%.
\citep{Moss+00}. The effect of the many further additional terms in the mean 
electromotive force \citep[e.g.][]{Rohde+99,Radler+03,Brandenburg+Subramanian05a} on the
pitch angle has never been explored.

%---------------------------------------------------------------------------------
\subsection{Arm-interarm contrast in magnetic field strength}
Various solutions of the dynamo equations offer a range of possibilities to explore
the effects of spiral arms on the large-scale galactic magnetic fields.
Among them are the phenomenon of magnetic arms, 
the enhancement of polarized intensity (and, presumably, 
the large-scale magnetic field) in spiral-shaped regions that do not always overlap with the gaseous arms, 
regions of larger gas density  \citep[e.g.,][]{Frick+00}. 
Several explanations have been suggested
(\citealt{Moss98,Shukurov98,Rohde+99,Chamandy+13a,Chamandy+13b,Moss+13}),
but there is no convincing explanation.

Equation \eqref{Bsat} suggests a number of effects that can contribute to a non-monotonic
dependence of the large-scale magnetic field strength on gas density.
One possibility is that $R_U$ can be larger in the arms owing 
to a greater frequency of supernova explosions there.
This can lead to larger $B^2/B\eq^2$ in the interarm regions compared to in the arms,
as suggested by \citet{Sur+07}.
An estimate of this effect can be found in Table~\ref{tab:no-z}:
for $r=8\kpc$ $\mean{B}$ would be about 1.5 times stronger in the interarm regions
in an extreme case where the outflow speed vanishes
between the arms remaining modest in the arms at $1\kms$ ($R_U=1.36$). 
This ratio increases to about 4 if $U\f=2\kms$ in the arms.
Therefore, this effect may be important and thus deserves a more detailed exploration.

%--------------------------------------------------------------------------------
\section{Conclusions and discussion}\label{sec:conc}\label{sec:summary}
We have discussed various simple approximate approaches to estimate the strength
of the large-scale galactic magnetic fields, their pitch angle and dynamo thresholds,
and compared them with numerical solutions. In particular, we compared the
non-linear states established due to magnetic helicity conservation
with those obtained with a much simpler, and easier to analyze, heuristic algebraic 
form of the dynamo non-linearity.
These approaches complement one another. 
For example the perturbation solution of Section~\ref{sec:pert} provides  
a reasonably accurate form of the distribution of magnetic field across the disc,
whereas the no-$z$ approximation of Section~\ref{sec:noz} gives useful
results for variables averaged across the disc.
Remarkably, and reassuringly, where they overlap, all of these methods result in similar solutions.
Most importantly, results obtained with the dynamical non-linearity that involves advective and
diffusive fluxes of magnetic helicity are very much consistent with those from
the algebraic $\alpha$-quenching. We suggest how the latter can be modified to achieve 
quite a detailed agreement.

Magnetic lines produced by the mean-field dynamo are believed to be trailing with 
respect to the galactic rotation because the galactic angular velocity decreases with 
galactocentric radius. We have found, however, that steady-state magnetic fields 
obtained for the dynamical non-linearity are trailing near the galactic midplane
but leading closer to the disc surface (where $\mean{B}_r$ changes sign) if an outflow is present. 
This effect is more pronounced when the galactic outflow is stronger
or the dynamo number is higher as compared with its critical value.
This feature is new and unexpected, as it is not reproduced in models with algebraic quenching.
This makes it reasonable to expect that leading magnetic spirals may be observable in the
disc-halo interface regions of spiral galaxies (or even higher in the halo).
To what extent this feature persists if the boundary conditions are varied or if the galactic halo is included
is a question that merits future investigation.

It is also useful for applications that marginal kinematic solutions of the
dynamo equations in a thin disc (i.e., those that neither grow nor decay) reproduce
with high accuracy non-linear steady-state solutions. The simple analytical
perturbation solutions of kinematic dynamo equations, here generalized to include 
magnetic field advection in a galactic outflow, are particularly useful in this
respect. It has been shown here and by \citet{Ji+13} that they remain accurate
beyond their formal range of applicability and can be used for the range of
dynamo numbers $-50\la D\la 0$ typical of galactic discs. Here we have also shown
that these solutions can be used as a good approximation to the non-linear states.

We have also refined the no-$z$ approximation to allow for 
vertical advection of the mean magnetic field,
as well as advective and diffusive helicity fluxes.
We note that advection affects dynamo action through three channels:
by reducing the critical dynamo number,
by helping the turbulent diffusion to remove flux from the dynamo active region,
and by the removal of small-scale magnetic helicity.
The heuristic diffusive flux of magnetic helicity has previously been observed in numerical simulations.

The models investigated here are somewhat simplified compared to real galaxies.
It is worth extending the models to include spatial variation of $\etat$,
additional terms in the mean electromotive force,
and other contributions to the magnetic helicity flux.
The possible importance of $\etat$-quenching, in addition to $\alpha$-quenching \citep{Gressel+13a},
also deserves exploration.
More refined modelling will enable better comparison with real galaxies.

In summary, much of the earlier work on galactic dynamos modeled the saturation of the dynamo 
using algebraic quenching of the $\alpha$ effect.
We show here that this algebraic quenching non-linearity
(which predates dynamical quenching theory but is still widely used in the dynamo literature)
is a good approximation to dynamical quenching for the galactic mean-field dynamo.
We also extend the standard algebraic quenching formula to make it more accurate.
In addition, we suggest three simple tools, namely marginal kinematic solutions,
critical asymptotic solutions from perturbation theory, 
and no-$z$ solutions, 
and show that all agree remarkably well with the numerical solutions of the non-linear dynamo.
Particularly useful are the analytical expressions \eqref{Brpert} and \eqref{Bppert} for the vertical profiles of $\mbr$ and $\mbp$,
as well as equations \eqref{Bsat} and \eqref{psat} for the saturation field strength $B$ and pitch angle $p$,
which, when used along with the analytical expression \eqref{gammapert} for the kinematic growth rate $\gamma$,
comprise a surprisingly efficient guide to the parameter space of galactic dynamos.

%-------------------------------------------------------------------------------
\section*{Acknowledgments}
We are grateful to Fred Gent and Aritra Basu for useful discussions,
and to Nishant Singh for critical reading of an earlier version of the manuscript.
AS acknowledges financial support of STFC (grant ST/L005549/1)
and the Leverhulme Trust (grant RPG-097),
and expresses his gratitude to IUCAA for hospitality and financial support.
%-------------------------------------------------------------------------------

\appendix
\section{Perturbation solutions of the dynamo equations with an outflow}
\label{sec:pert_app}

The kinematic $\alpha\Omega$ dynamo in a thin disc is governed by equation \eqref{mbr} and equation \eqref{mbp} 
with the $\alpha$ term omitted in equation \eqref{mbp}.
These can be written in dimensionless form as \citep[e.g.][]{Shukurov04}:
\begin{align}
  \label{mbrdim}
  \frac{\del\mbr}{\del t}&=-R_\alpha\frac{\del}{\del z}(\alphatilde\mbp) +\frac{\del^2\mbr}{\del z^2} -R_U\frac{\del}{\del z}(\widetilde{U}_z\mbr),\\
  \label{mbpdim}
  \frac{\del\mbp}{\del t}&= R_\omega\mbr +\frac{\del^2\mbp}{\del z^2} -R_U\frac{\del}{\del z}(\widetilde{U}_z\mbp),
\end{align}
with the vacuum boundary conditions at the disc surface,
\begin{equation*}
  \mbr|_{z=\pm1}=\mbp|_{z=\pm1}=0,
\end{equation*}
and where $\mbz$ can be recovered from the solenoidality condition.
We make the equations time-independent by substituting the solution $\meanv{B}=\bm{\Bs}(z)\Exp{\gamma t}$,
and make the transformation 
\begin{equation}
  \label{Btransf}
  \Bsr'\equiv R_\alpha^{-1}\Bsr, \quad \Bsp'\equiv \Bsp/\sqrt{|D|}, 
\end{equation}
where $D\equiv R_\omega R_\alpha$,
and then drop primes for presentational convenience,
so that equations \eqref{mbrdim} and \eqref{mbpdim} become
\begin{equation*}
  \begin{split}
    \gamma\Bsr&= -\sqrt{|D|}\frac{d}{d z}(\alphatilde\Bsp) +\frac{d^2\Bsr}{d z^2} -R_U\frac{d}{d z}(\widetilde{U}_z\Bsr),\\
    \gamma\Bsp&= \sqrt{|D|}\mathrm{sign}(D)\Bsr +\frac{d^2\Bsp}{d z^2} -R_U\frac{d}{d z}(\widetilde{U}_z\Bsp).
  \end{split}
\end{equation*}
We then seek an asymptotic solution to this eigenvalue problem by treating terms involving $\sqrt{|D|}$ and $R_U$ 
as a perturbation to the eigensolutions of the `free-decay' modes ($D=R_U=0$) of the diffusion equation.
This generalizes the treatment of \citet{Shukurov04,Sur+07,Shukurov+Sokoloff08}, 
who solve the case $R_U=0$.
We may write
\begin{equation}
  \label{pert}
  (\widehat{W} +\epsilon\widehat{V})\bm{\Bs}= \gamma\bm{\Bs},
\end{equation}
where $\epsilon=\const$ is a mathematical device to keep track of the orders, 
and is taken to be $\ll1$ for the perturbation analysis before being `restored' to its true value of unity at the end of the calculation 
\citep[e.g.][]{Griffiths05}.
The unperturbed operator $\widehat{W}$ is given by
\begin{eqnarray*}
  \widehat{W}=\left(
    \begin{array}{cc}
      \displaystyle\frac{d^2}{dz^2} & 0                            \\
      0                             & \displaystyle\frac{d^2}{dz^2}
    \end{array}
  \right),
\end{eqnarray*}
while the perturbation operator $\widehat{V}$ is given by
\begin{eqnarray*}
  \widehat{V}=\left(
    \begin{array}{cc}
      -\displaystyle R_U\frac{d}{dz}(z\cdots) & -\displaystyle \sqrt{-D}\frac{d}{dz}\left[\sin(\pi z)\cdots\right]\\
      -\sqrt{-D}                              & -\displaystyle R_U\frac{d}{dz}(z\cdots)
    \end{array} 
  \right),
\end{eqnarray*} 
where we have adopted the forms \eqref{Krause} and \eqref{muztilde} for $\alphatilde$ and $\widetilde{U}_z$,
and taken $\mathrm{sign}(D)=-1$, as is suitable for the present context.

Keeping terms containing both $\sqrt{-D}$ and $R_U$,
we effectively assume that they are of the same order of magnitude.
However, \citet{Ji+13} show that the resulting perturbation solution remains accurate up to $\sqrt{-D}\gtrsim1$
(they consider the case $R_U=0$).
Then the requirement that $\sqrt{-D}=\mathcal{O}(R_U)$ is rather formal and not restrictive for practical purposes.

The eigensolutions of the unperturbed system $\What\bmb_n= \lambda_n\bmb_n$ (with the above boundary conditions)
are doubly degenerate and given by
\begin{equation*}
  \label{lambda_n}
  \lambda_n=-\left( n +\tfrac{1}{2}\right)^2\pi^2,\quad n= 0, 1, 2, \ldots,
\end{equation*}

\begin{equation*}
  \bmb_{n}=\left(
    \begin{array}{c}
      \sqrt2\cos[(n +\tfrac{1}{2})\pi z]\\
      0
    \end{array}
  \right),
\end{equation*}

\begin{equation*}
  \bmb'_{n}=\left(
    \begin{array}{c}
      0                               \\
      \sqrt2\cos[(n +\tfrac{1}{2})\pi z]
    \end{array}
  \right),
\end{equation*}
where eigenfunctions have been normalized to $\int^1_0\bmb_n^2dz= \int^1_0{\bmb'_n}^2dz= 1$
(the eigenfunctions should not be confused with the small-scale magnetic field,
denoted $\bmb$ in the main text).

\begin{figure}                     
  \includegraphics[width=\columnwidth,clip=true,trim= 0 10 0 0]{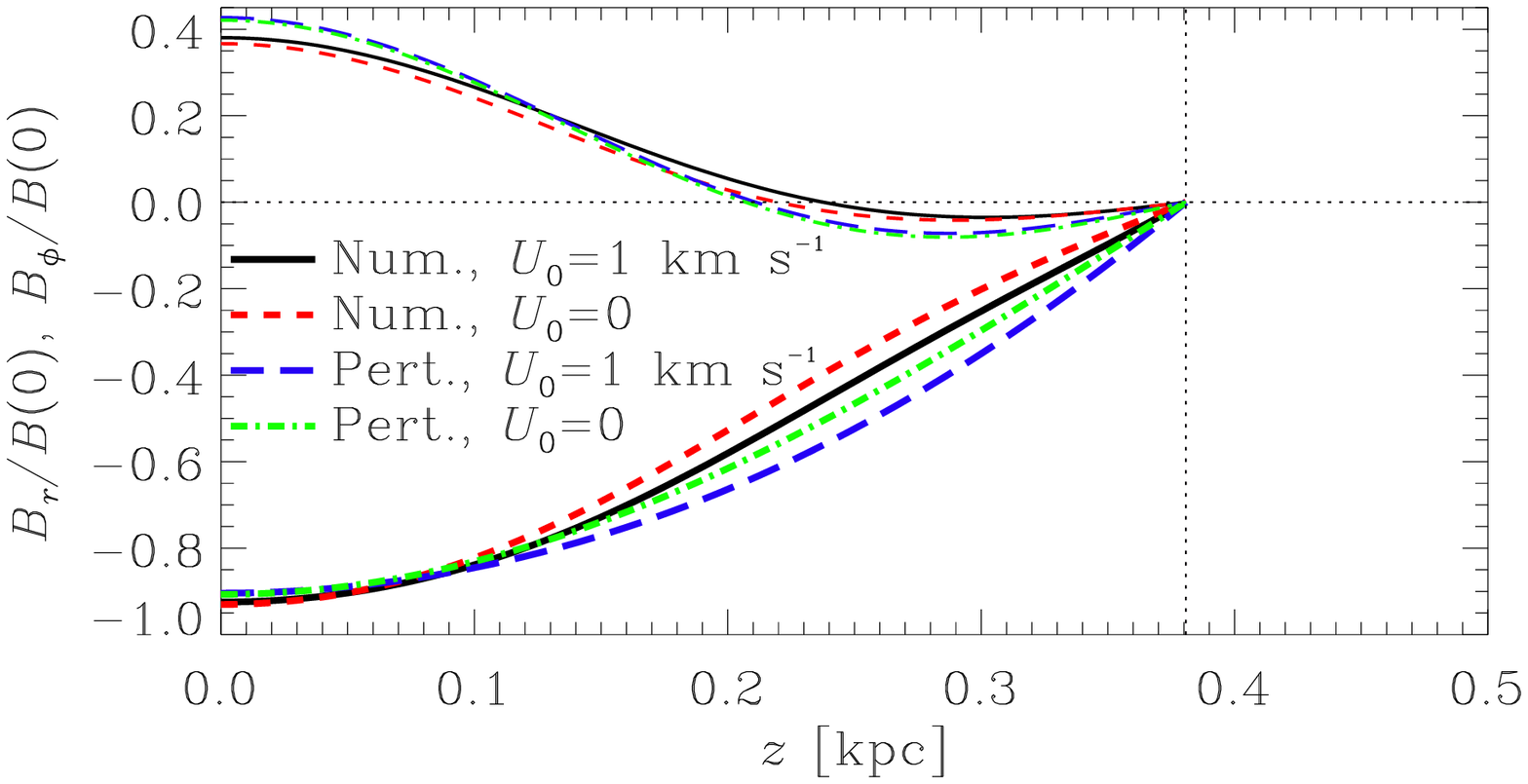}\\
  \vspace{-0.20cm}
  \includegraphics[width=\columnwidth,clip=true,trim= 0 0 0 10]{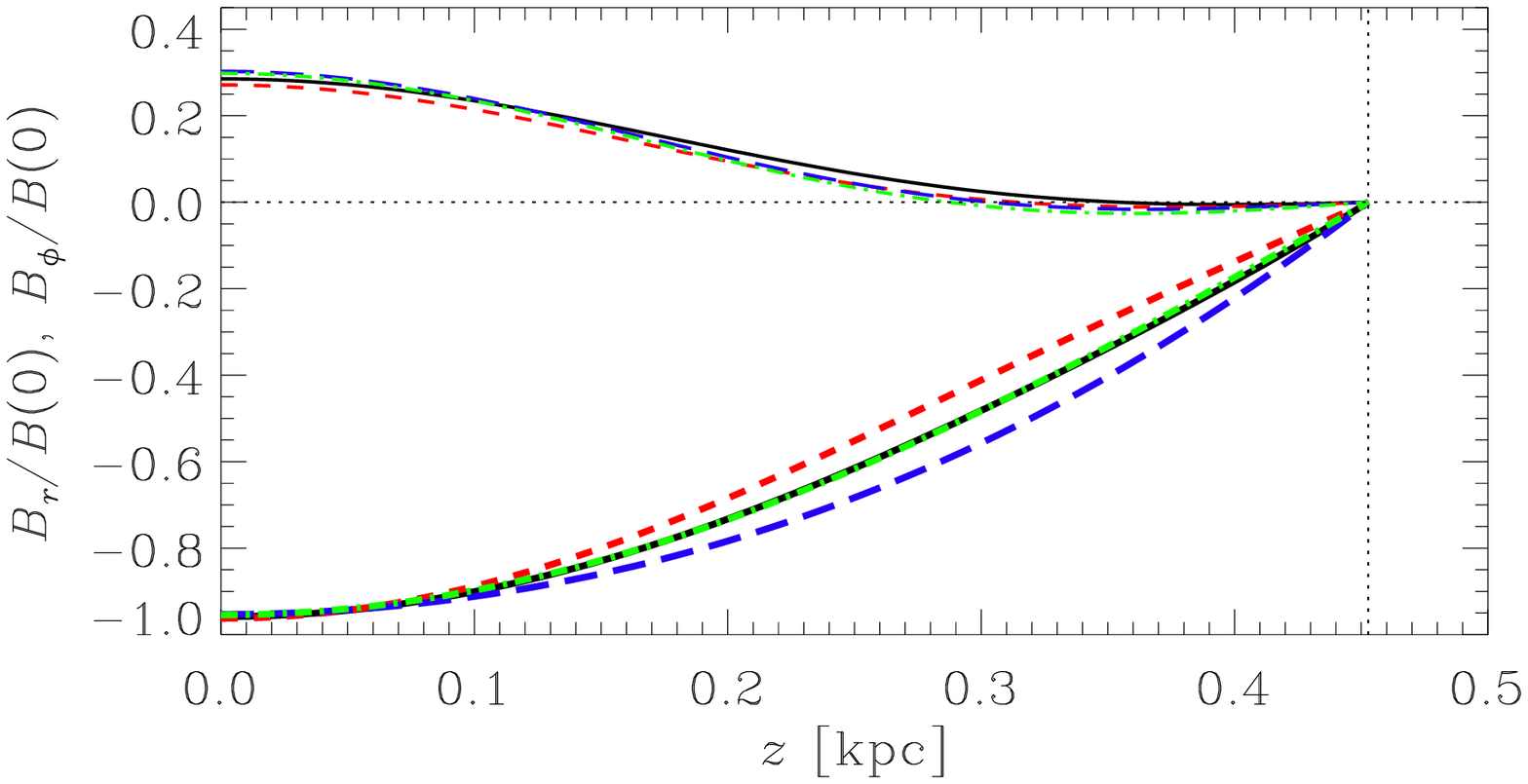}
  \caption{Radial (thin) and azimuthal (thick) components of $\meanv{B}$ in the kinematic stage, 
           normalized to the magnetic field strength at the midplane
           for parameters corresponding to $r=4\kpc$ (top) and $r=8\kpc$ (bottom).
           Solutions from perturbation theory [equations \eqref{Brpert_kin} and \eqref{Bppert_kin}]
           are compared with numerical solutions for $U\f=0$ and $U\f=1\kms$.
           Solutions are symmetric about the midplane.
           \label{fig:Bvsz_kin}
          }            
\end{figure}

\begin{figure}                     
  \includegraphics[width=\columnwidth,clip=true,trim= 0 10 0 0]{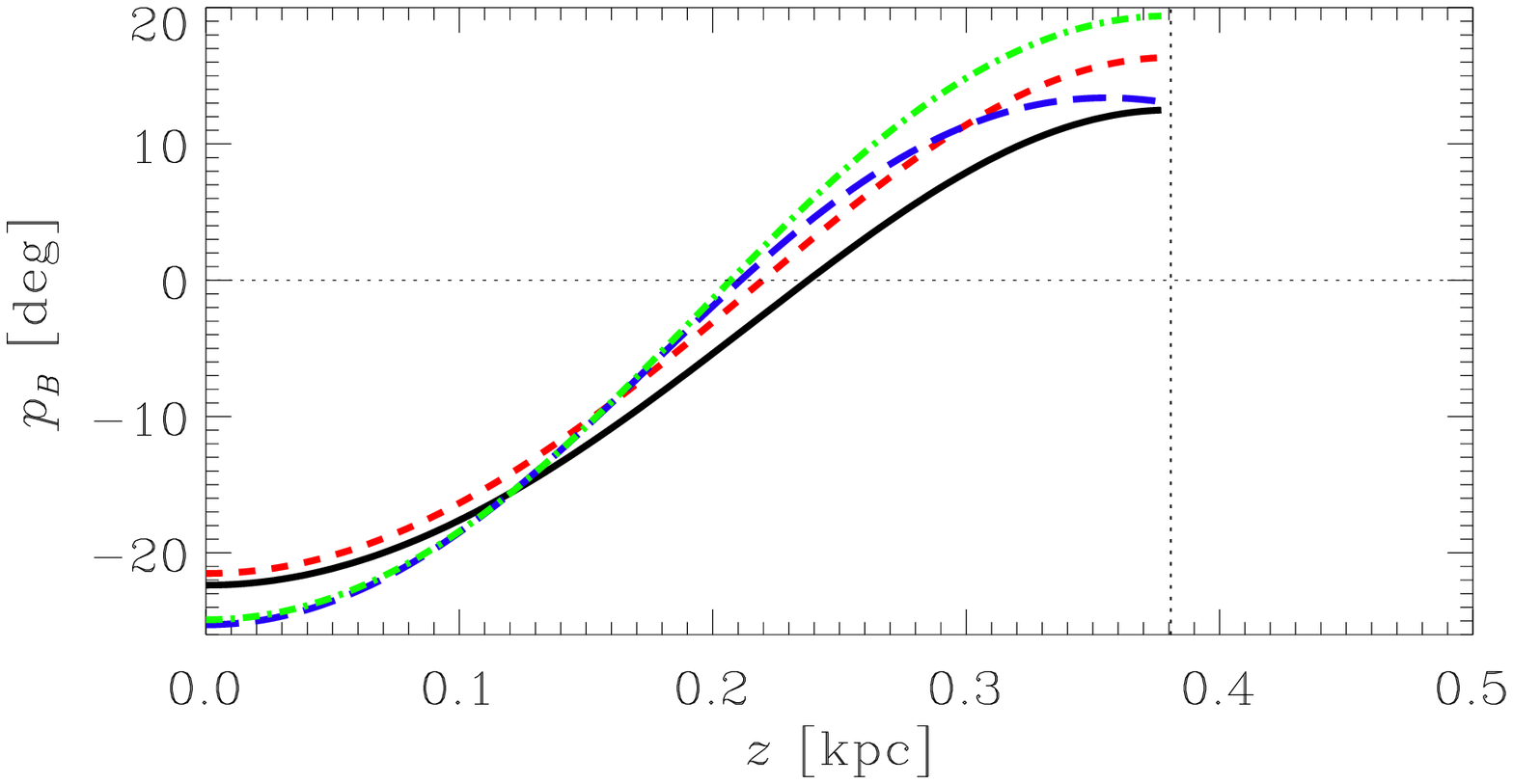}\\
  \vspace{-0.20cm}
  \includegraphics[width=\columnwidth,clip=true,trim= 0 0 0 10]{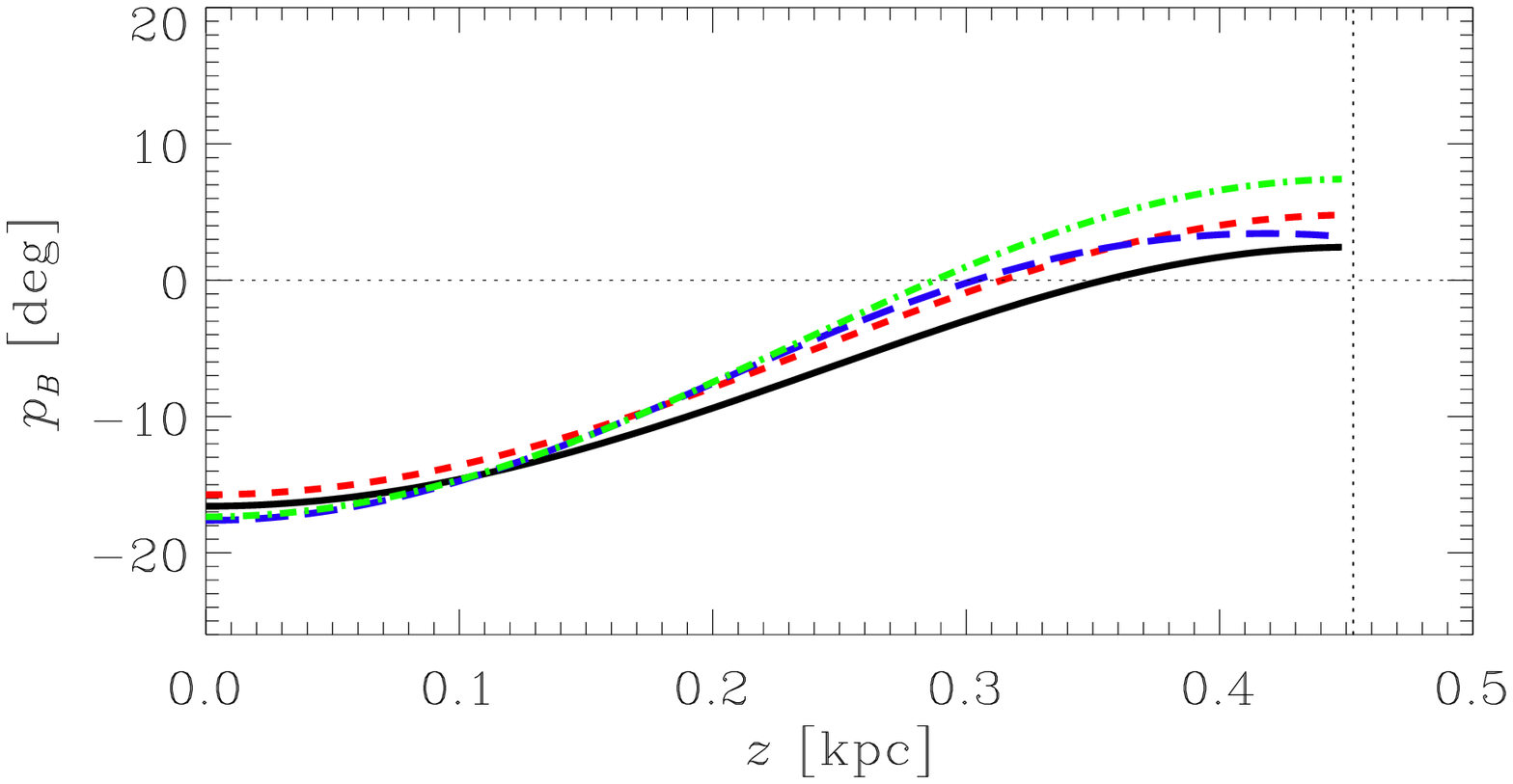}
  \caption{Magnetic pitch angle $p\equiv\arctan(\mbr/\mbp)$ in the kinematic stage, 
           as a function of the distance $z$ from the midplane,
           for parameters corresponding to $r=4\kpc$ (top) and $r=8\kpc$ (bottom).
           $p$ is not plotted for $z=h$, as it is undefined at the disc boundaries, 
           where the boundary conditions enforce $\mbr=\mbp=0$.
           (See also Fig.~\ref{fig:Bvsz_kin}.)
           \label{fig:pBvsz_kin}
          }            
\end{figure}

The expansions
\begin{equation*}
  \gamma= \gamma\f +\epsilon\gamma_1 +\epsilon^2\gamma_2 +\ldots,
\end{equation*}
\begin{equation*}
  \label{eigenfn_expansion}
  \bm{\Bs}= C\f\bmb\f +C'\f\bmb'\f +\epsilon\displaystyle\sum^\infty_{n=1}(C_n\bmb_n +C'_n\bmb'_n)+ \ldots
\end{equation*}
are substituted into equation \eqref{pert}, terms of like order in $\epsilon$ collected,
the dot product of the resulting equations taken first with $\bmb_n$ and then with $\bmb'_n$,
and the results integrated over $0\le z\le 1$.
[Note that the lowest order contribution to the eigenfunction 
can be assumed to be a linear combination of the fastest growing ($n=0$) terms only.]
To the lowest order this yields $\gamma\f= \lambda\f$.
A homogeneous system of algebraic equations for $C\f$ and $C'\f$ follows from terms of order $\epsilon$,
whose solvability condition yields 
\begin{equation*}
  \begin{split}
    \gamma_1&= \frac{1}{2}\left\{(V_{00} +V_{0'0'}) + \left[ (V_{00} -V_{0'0'})^2 +4V_{00'}V_{0'0}\right]^{1/2}\right\}\\
            &= -\frac{R_U}{2} +\frac{\sqrt{-\pi D}}{2},
  \end{split}
\end{equation*}
and 
\begin{equation*}
  C'\f= \frac{(\gamma_1 -V_{00})}{V_{00'}}C\f= -\frac{2}{\sqrt{\pi}}C\f,
\end{equation*}
where we have retained only the root corresponding to the growing solution.
Here $V_{nm}\equiv\int^1_0\bmb_n\cdot\Vhat\bmb_m dz$ are the perturbation matrix elements,
whose direct calculation yields 
\begin{equation*}
  \begin{split}
    V_{00'}= -\pi\sqrt{-D}/4,\quad V_{0'0}= -\sqrt{-D},\\
    V_{00}=V_{0'0'}= -\frac{R_U}{2}.
  \end{split}
\end{equation*}
The eigenvalue can be evaluated to a higher order in $\epsilon$ than the eigenfunction
since it depends on the matrix elements $V_{\tilde{0}n}$, $V_{\tilde{0}n'}$, $V_{n\tilde{0}}$, 
and $V_{n'\tilde{0}}$, where, e.g., $V_{\tilde{0}n}\equiv\int^1_0\tilde{\bmb}_0\cdot\Vhat\bmb_n dz$,
and 
\begin{equation*}
  \tilde{\bmb}\equiv C\f\bmb\f +C'\f\bmb'\f= C\f\left(\begin{array}{c}1\\-2/\sqrt{\pi}\end{array}\right)\sqrt{2}\cos\left(\frac{\pi z}{2}\right).
\end{equation*}
The above method then yields:
\begin{equation*}
  \gamma_2= \displaystyle\sum^\infty_{n=1}\frac{V_{n\tilde{0}}V_{\tilde{0}n} +V_{n'\tilde{0}}V_{\tilde{0}n'}}{\lambda\f -\lambda_n}, 
\end{equation*}
and
\begin{equation*}
  \quad C_n= \frac{V_{n\tilde{0}}}{\lambda\f -\lambda_n}, \quad C'_n= \frac{V_{n'\tilde{0}}}{\lambda\f - \lambda_n},
\end{equation*}
where
\begin{eqnarray*}
  &V_{n\tilde{0}}&= C\f\times\left\{
    \begin{array}{cc}
      \displaystyle\frac{3\sqrt{-\pi D}}{2}-\frac{3R_U}{4}, & n=1;\\[3mm]
      \displaystyle\frac{2n+1}{n(n+1)}\frac{(-1)^{n}R_U}{2}, & n\geq 2.
    \end{array}
  \right.\\
  \vspace{2mm}
  &V_{n'\tilde{0}}&=C\f\times\left\{
    \begin{array}{cc}
      \displaystyle-\frac{2n+1}{n(n+1)}\frac{(-1)^{n}R_U}{\sqrt{\pi}}, & n\geq 1.
    \end{array}
  \right.\\
  \vspace{2mm}
  &V_{\tilde{0}n}&=C\f\times\left\{
    \begin{array}{cc}
      \displaystyle-\frac{2n+1}{n(n+1)}\frac{(-1)^{n}R_U}{2}, & n\geq 1.
    \end{array}
  \right.\\
  \vspace{2mm}
  &V_{\tilde{0}n'}&=C\f\times\left\{
    \begin{array}{cc}
      \displaystyle\frac{\pi\sqrt{-D}}{4}-\frac{3R_U}{2\sqrt{\pi}}, & n=1;\\[3mm]
      \displaystyle\frac{2n+1}{n(n+1)}\frac{(-1)^{n}R_U}{\sqrt{\pi}}, & n\geq 2.
  \end{array}\right.
\end{eqnarray*}
Using the fact that
\begin{equation*}
  \displaystyle\sum^\infty_{n=2}\frac{(2n+1)^2}{2n^3(n+1)^3}= -\frac{25}{16} +\frac{\pi^2}{6},
\end{equation*}
we find
\begin{equation*}
  \gamma_2= \frac{3R_U}{4\sqrt{\pi}(\pi+4)} +\frac{R_U^2}{2\pi^2}\left(1 -\frac{\pi^2}{6}\right).
\end{equation*}
For the eigenfunction, we obtain
\begin{equation*}
  \begin{split}
  &\displaystyle\sum^\infty_{n=1}(C_n\bmb_n + C'_n\bmb'_n)= \\
  &\quad\frac{C\f}{\pi^2}\Bigg\{\frac{3}{4}\left[ \sqrt{-\pi D}\bmb_1 -\frac{R_U}{2}\left(\bmb_1 -\frac{2\bmb'_1}{\sqrt{\pi}}\right)\right]\\
  &\quad\left.+\frac{R_U}{2}\displaystyle\sum^\infty_{n=2}\left[\frac{(-1)^n(2n+1)}{n^2(n+1)^2}\left(\bmb_n -\frac{2\bmb'_n}{\sqrt{\pi}}\right)\right]\right\}.
  \end{split}
\end{equation*}
Thus, the final solution of second order perturbation theory, upon restoring $\epsilon=1$, 
and also restoring the original definitions of $\Bs_r$ and $\Bs_\phi$ in equation \eqref{Btransf},
reduces to
\begin{equation*}
    \gamma=-\frac{\pi^2}{4} +\frac{\sqrt{-\pi D}}{2} -\frac{R_U}{2} +\frac{3\sqrt{-\pi D}R_U}{4\pi(\pi+4)}
           +\frac{R_U^2}{2\pi^2}\left(1 -\frac{\pi^2}{6}\right),
\end{equation*}
\begin{equation}\label{Brpert_kin}
  \begin{split}
    \Bs_r= &C\f R_\alpha\Bigg\{\cos\left(\frac{\pi z}{2}\right) +\frac{3}{4\pi^2}\left( \sqrt{-\pi D} 
            -\frac{R_U}{2}\right)\cos\left(\frac{3\pi z}{2}\right)\\
           &\quad\left.+\frac{R_U}{2\pi^2}\displaystyle\sum^\infty_{n=2}\frac{(-1)^n(2n+1)}{n^2(n+1)^2}
            \cos\left[\left(n+\tfrac{1}{2}\right)\pi z\right]\right\},
  \end{split}
\end{equation}
\begin{equation}\label{Bppert_kin}
  \begin{split}
    \Bs_\phi= &-\frac{2}{\pi}C\f\sqrt{-\pi D}\Bigg\{ \cos\left(\frac{\pi z}{2}\right) 
               -\frac{3R_U}{8\pi^2}\cos\left(\frac{3\pi z}{2}\right)\\
              &\quad\left.+\frac{R_U}{2\pi^2}\displaystyle\sum^\infty_{n=2}\frac{(-1)^n(2n+1)}{n^2(n+1)^2}
               \cos\left[\left(n+\tfrac{1}{2}\right)\pi z\right]\right\},
  \end{split}
\end{equation}
where $C\f=1/\sqrt{1+4/\pi}$ for the solution normalized as $\int^1_0 B^2dz=1$.
The eigenfunctions are plotted in Fig.~\ref{fig:Bvsz_kin} and the magnetic pitch angle, in Fig.~\ref{fig:pBvsz_kin}.
Solving for the critical ($\gamma=0$) dynamo number we obtain
\begin{equation*}
  D\crit=-\frac{\pi^3}{4}\left\{ \frac{ 1 +2R_U/\pi^2 -( 2R_U^2/\pi^4)( 1 -\pi^2/6)}{ 1 +3R_U/[ 2\pi( \pi +4)]}\right\}^2.
\end{equation*}

%---------------------------------------------------------------------------------

%---------------------------------------------------------------------------------
\section{The no-\large{\MakeLowercase{$z$}} asymptotic solution}
\label{sec:noz_app}

Equations \eqref{mbr}-\eqref{alpha_m} can be solved in an approximate way as a set of algebraic equations by setting time derivatives to zero
(that is, by assuming the system reaches a steady state)
and by using the no-$z$ approximation to replace $z$-derivatives by simple divisions.
This method allows for a determination of all relevant quantities, 
but these quantities now represent averages over the disc half-thickness $h$.
The method is not new \citep{Sur+07}, but here we neglect Ohmic terms, include the diffusive flux,
and do not assume that $\mbr^2/\mbp^2\ll1$.
We do, however, adopt the $\alpha\Omega$ approximation for simplicity.
We also include an expression for the growth rate $\gamma$ in the kinematic regime, 
which is obtained by assuming $\meanv{B}\propto\Exp{\gamma t}$.

\begin{figure*}                     
  %\hspace{-1.0cm}
  \includegraphics[height=3.7cm,clip=true,trim= 70 40 20 30]{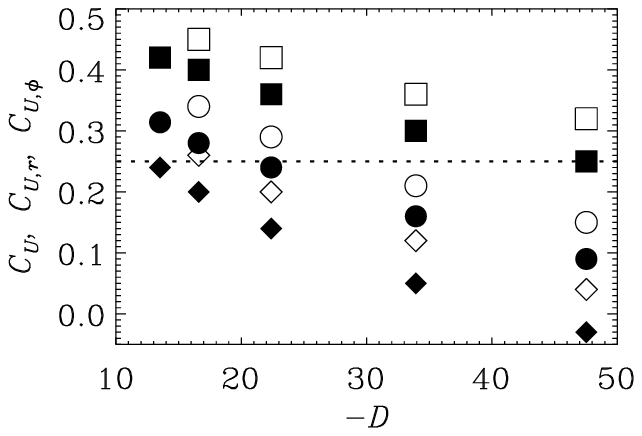}
  %\hspace{-1.0cm}
  \includegraphics[height=3.7cm,clip=true,trim= 70 40 20 30]{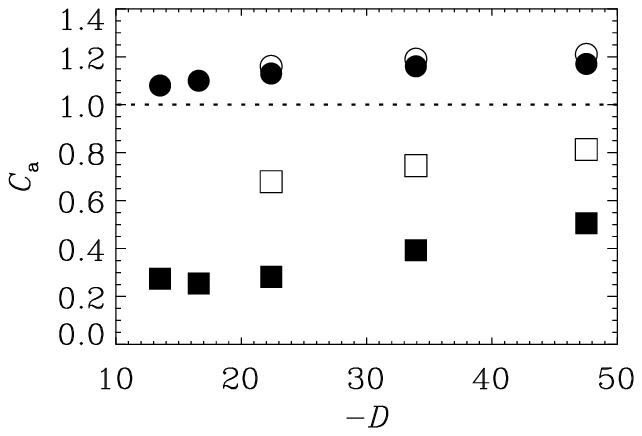}
  %\hspace{-1.0cm}
  \includegraphics[height=3.7cm,clip=true,trim= 55 40 20 30]{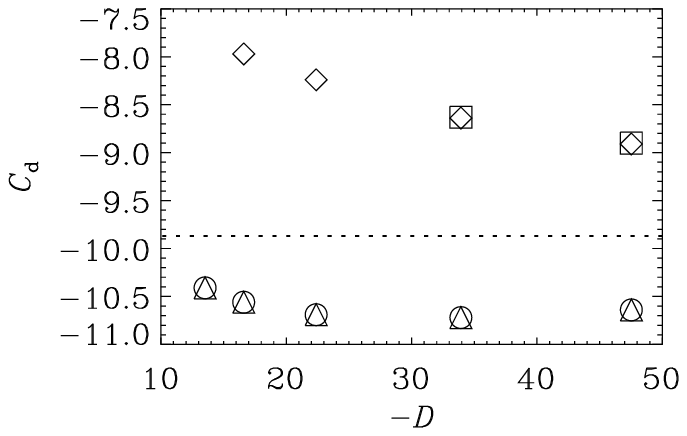}
  \caption{Left:   Parameter $C_U$ for $U\f=1\kms$ (filled circles) and $2\kms$ (open circles),
                   $C_{U,r}$ for $U\f=1\kms$ (filled diamonds) and $2\kms$ (open diamonds),
                   $C_{U,\phi}$ for $U\f=1\kms$ (filled squares) and $2\kms$ (open squares).
           Middle: Parameter $C\adv$ for $U\f=1\kms$ and $R_\kappa=0$ (filled circles),
                   $U\f=2\kms$ and $R_\kappa=0$ (open circles),
                   $U\f=1\kms$ and $R_\kappa=0.3$ (filled squares),
                   and $U\f=2\kms$ and $R_\kappa=0.3$ (open squares).
           Right:  Parameter $C\diff$ for $R_\kappa=0.3$ and $U\f=0$ (circles),
                   $R_\kappa=0.3$ and $U\f=1\kms$ (diamonds),
                   $R_\kappa=0.3$ and $U\f=2\kms$ (squares),
                   and $R_\kappa=0.6$ and $U\f=0$ (triangles).
                   Data for five different dynamo numbers $D=-47.5,-33.9,-22.4,-16.6,-13.5$,
                   corresponding to radii $r=2,4,6,8,10\kpc$, respectively,
                   is shown. 
                   Chosen values $C_U=0.25$, $C\adv=1$ and $C\diff=-\pi^2$
                   are shown by dashed lines on their respective graphs.
           \label{fig:nozacc}
          }            
\end{figure*}

Furthermore, we include as yet unspecified numerical factors in the no-$z$ terms.
There are in general four such factors that are not already specified in \citet{Phillips01}: 
$C_{U,r}$, such that $\del(\muz\mbr)/\del z\simeq C_{U,r}\muz\mbr/h$,
$C_{U,\phi}$, where, similarly, $\del(\muz\mbp)/\del z\simeq C_{U,\phi}\muz\mbp/h$,
$C\adv$, where $\del(\muz\alpha\magn)/\del z\simeq C\adv\muz\alpha\magn/h$ for the advective flux term,
and finally $C\diff$, with $\del^2\alpha\magn/\del z^2\simeq C\diff\alpha\magn/h^2$ for the diffusive flux term.
For simplicity, we approximate $C_{U,r}=C_{U,\phi}=C_U$, 
which turns out to be fairly reasonable (see Fig.~\ref{fig:nozacc} and the discussion in Section~\ref{sec:nozacc}).

Equations \eqref{mbrdim} and \eqref{mbpdim} can be written in the no-$z$ approximation, 
in dimensionless form ($h=\etat=1$) as
\begin{align*}
  \left(\gamma+\frac{\pi^2}{4}g\right)\mbr&= -\frac{2}{\pi}R_\alpha\mbp,\\
  \left(\gamma+\frac{\pi^2}{4}g\right)\mbp&= R_\omega\mbr,
\end{align*}
where $g\equiv 1+4C_UR_U/\pi^2$.
From the solvability condition for the homogeneous equations $(\gamma+\pi^2g/4)^2=-2D/\pi$,
we set $\gamma=0$ to obtain the critical dynamo number,
\begin{equation*}
  D\crit= -\frac{\pi^5}{32}g^2.
\end{equation*}
Solving for $\gamma$, we then obtain
\begin{equation*}
  \gamma= \frac{\pi^2}{4}t\diff^{-1}g\left(\sqrt{\frac{D}{D\crit}} -1\right),
\end{equation*}
where $t\diff= h^2/\etat$ is the turbulent diffusion time-scale.
Defining $p\equiv \arctan(\mbr/\mbp)$, 
and letting $R_\alpha=R_{\alpha,\mathrm{c}}$ since we are interested in the saturated solution,
we obtain
\begin{equation}
  \label{tanp}
  \tan p= \sqrt{-\frac{2R_{\alpha,\mathrm{c}}}{\pi R_\omega}}= \frac{\pi^2}{4}\frac{g}{R_\omega}.
\end{equation}
Finally, for the saturation field strength we use equation \eqref{alpha_m} with the left hand side equal to zero,
the expression
\begin{equation}
\label{jdotb}
  (\bmDel\cro\meanv{B})\cdot\meanv{B}\simeq -\frac{3\sqrt{-\pi D\crit}\tan (p) B^2}{8h[1+\tan^2(p)]}
\end{equation}
valid in the no-$z$ approximation \citep{Sur+07},
along with the above expression \eqref{tanp} for $\tan p$ in terms of $R_{\alpha,\mathrm{c}}$.
It is then straightforward to obtain
\begin{equation*}
  B^2= B\eq^2\frac{\xi(p)}{C}\left(\frac{D}{D\crit} -1\right)\left(C\adv R_U -C\diff R_\kappa\right),
\end{equation*}
where $\xi(p)\equiv[1-3\cos^2(p)/(4\sqrt{2})]^{-1}$ and $C\equiv 2(h/l)^2$.
We note that $C\diff<0$ (Section~\ref{sec:nozacc}).

%--------------------------------------------------------------------------------------

%-------------------------------------------------------------------------------------
\subsection{Refining the no-$z$ approximation}\label{sec:nozacc}
The values of $C_U$, $C\adv$ and $C\diff$ are estimated in the following way. 
To estimate $C_U$, 
equations \eqref{mbr} and \eqref{mbp} are first solved numerically 
and the true value of the growth rate $\gamma$ obtained.
The no-$z$ approximation is then applied \textit{to only those terms involving $C_U$}, 
e.g. $\del(\muz\mbr)/\del z$ is replaced with $C_U U\f\mbr/h$ (but $\mbr$ is still $z$-dependent).
The value of $C_U$ is then varied iteratively until $\gamma$ is equal to the true value.
This is also done for $C_{U,r}$ and $C_{U,\phi}$ 
by using the method on each of the relevant terms, individually.
To estimate $C\adv$ and $C\diff$, the same approach is taken, 
with the relevant term involving $C\adv$ or $C\diff$ replaced with its no-$z$ form,
but now equations \eqref{mbr}-\eqref{alpha_m} are solved,
and instead of matching the growth rate, 
\begin{equation*}
\label{<b>}
  \langle B \rangle\equiv \frac{1}{2h}\int^h_{-h}B dz,
\end{equation*}
is matched to its true value.
Alternatively, $\langle B^2 \rangle^{1/2}$ could be chosen, 
but it was found that this choice leads to very similar results.

The results are summarized in Fig.~\ref{fig:nozacc}.
Unfortunately, $C_U$, represented by circles in the left panel,
has a fairly strong dependence on $D$, 
but luckily, a rather weak dependence on $R_U$.
We adopt the value $C_U=0.25$, which seems to be a reasonable choice given the data.
The middle panel shows the values of $C\adv$ obtained for different $D$, $R_U$ and $R_\kappa$.
Values are close to unity for $R_\kappa=0$,
but drop quite drastically when $R_\kappa=0.3$.
We adopt $C\adv=1$, 
keeping in mind that for cases with both advective and diffusive flux,
this is an overestimate.
Finally, the results for $C\diff$ are illustrated in the right panel of Fig.~\ref{fig:nozacc}.
Its value does not stray very far from $-\pi^2$, 
which is the value that would be obtained if $\alpha\magn$ were sinusoidal in $z$.

We therefore adopt the values $C_U=1/4$, $C\adv=1$, and $C\diff=-\pi^2$ in the present work.
It is worth mentioning that these choices may not be as suitable for a different $\muz$ profile.
With these choices, 
approximating a given term by its no-$z$ form 
leads to errors in quantities such as $\gamma$, $\langle B/B\eq\rangle$ or the pitch angle $p$ 
of typically $<10\%$ for parameters corresponding to $r=4\kpc$ or $r=8\kpc$ in our model. 
We fully realize that the values of these numerical factors,
and the method used to determine them, 
are not exact or unique; 
the idea is to improve the no-$z$ approximation in the same spirit as \citet{Phillips01},
while admitting that the approximation itself is rather crude by nature.

%-----------------------------------------------------------------------------------------------------------------
\bibliographystyle{mn2e}
\bibliography{refs}

\begin{thebibliography}{53}
\expandafter\ifx\csname natexlab\endcsname\relax\def\natexlab#1{#1}\fi

\bibitem[{{Bardou} {et~al.}(2001){Bardou}, {von Rekowski}, {Dobler},
  {Brandenburg}, \& {Shukurov}}]{Bardou+01}
{Bardou} A., {von Rekowski} B., {Dobler} W., {Brandenburg} A., {Shukurov} A.,
  2001, \aap, 370, 635

\bibitem[{{Beck}(2007)}]{Beck07}
{Beck} R., 2007, \aap, 470, 539

\bibitem[{{Beck} {et~al.}(1996){Beck}, {Brandenburg}, {Moss}, {Shukurov}, \&
  {Sokoloff}}]{Beck+96}
{Beck} R., {Brandenburg} A., {Moss} D., {Shukurov} A., {Sokoloff} D., 1996,
  \araa, 34, 155

\bibitem[{{Bernet} {et~al.}(2013){Bernet}, {Miniati}, \& {Lilly}}]{Bernet+13}
{Bernet} M.~L., {Miniati} F., {Lilly} S.~J., 2013, \apjl, 772, L28

\bibitem[{{Blackman}(2014)}]{Blackman14}
{Blackman} E.~G., 2014, ArXiv e-prints

\bibitem[{{Blackman} \& {Field}(2000)}]{Blackman+Field00}
{Blackman} E.~G., {Field} G.~B., 2000, \apj, 534, 984

\bibitem[{{Brandenburg}(2003)}]{Brandenburg03}
{Brandenburg} A., 2003, {Computational aspects of astrophysical MHD and
  turbulence}, Taylor \& Francis, London, p. 269

\bibitem[{{Brandenburg} {et~al.}(2009){Brandenburg}, {Candelaresi}, \&
  {Chatterjee}}]{Brandenburg+09}
{Brandenburg} A., {Candelaresi} S., {Chatterjee} P., 2009, \mnras, 398, 1414

\bibitem[{{Brandenburg} \& {Subramanian}(2005)}]{Brandenburg+Subramanian05a}
{Brandenburg} A., {Subramanian} K., 2005, \physrep, 417, 1

\bibitem[{{Chamandy} {et~al.}(2014){Chamandy}, {Subramanian}, \&
  {Quillen}}]{Chamandy+13d}
{Chamandy} L., {Subramanian} K., {Quillen} A., 2014, \mnras, 437, 562

\bibitem[{{Chamandy} {et~al.}(2013{\natexlab{a}}){Chamandy}, {Subramanian}, \&
  {Shukurov}}]{Chamandy+13a}
{Chamandy} L., {Subramanian} K., {Shukurov} A., 2013{\natexlab{a}}, \mnras,
  428, 3569

\bibitem[{{Chamandy} {et~al.}(2013{\natexlab{b}}){Chamandy}, {Subramanian}, \&
  {Shukurov}}]{Chamandy+13b}
---, 2013{\natexlab{b}}, \mnras, 433, 3274

\bibitem[{{Chy{\.z}y} \& {Buta}(2008)}]{Chyzy+Buta08}
{Chy{\.z}y} K.~T., {Buta} R.~J., 2008, \apjl, 677, L17

\bibitem[{{Ebrahimi} \& {Bhattacharjee}(2014)}]{Ebrahimi+Bhattacharjee14}
{Ebrahimi} F., {Bhattacharjee} A., 2014, ArXiv e-prints

\bibitem[{{Elstner}(2005)}]{Elstner05}
{Elstner} D., 2005, in The Magnetized Plasma in Galaxy Evolution, {Chyzy}
  K.~T., {Otmianowska-Mazur} K., {Soida} M., {Dettmar} R.-J., eds.,
  Jagiellonian University, Krakow, pp. 117--124

\bibitem[{{Fletcher}(2010)}]{Fletcher10}
{Fletcher} A., 2010, in Astron Soc Pacific Conf Ser, {Kothes} R., {Landecker}
  T.~L., {Willis} A.~G., eds., Vol. 438, p. 197

\bibitem[{{Frick} {et~al.}(2000){Frick}, {Beck}, {Shukurov}, {Sokoloff},
  {Ehle}, \& {Kamphuis}}]{Frick+00}
{Frick} P., {Beck} R., {Shukurov} A., {Sokoloff} D., {Ehle} M., {Kamphuis} J.,
  2000, \mnras, 318, 925

\bibitem[{{Gressel} {et~al.}(2013){Gressel}, {Bendre}, \&
  {Elstner}}]{Gressel+13a}
{Gressel} O., {Bendre} A., {Elstner} D., 2013, \mnras, 429, 967

\bibitem[{Griffiths(2005)}]{Griffiths05}
Griffiths D., 2005, Introduction to Quantum Mechanics. Pearson Education, Essex

\bibitem[{{Gruzinov} \& {Diamond}(1994)}]{Gruzinov+Diamond94}
{Gruzinov} A.~V., {Diamond} P.~H., 1994, Phys Rev Lett, 72, 1651

\bibitem[{{Heald}(2012)}]{Heald12}
{Heald} G.~H., 2012, \apjl, 754, L35

\bibitem[{{Ji} {et~al.}(2013){Ji}, {Cole}, {Bushby}, \& {Shukurov}}]{Ji+13}
{Ji} Y., {Cole} L., {Bushby} P., {Shukurov} A., 2013, ArXiv e-prints 1312.0408

\bibitem[{{Kleeorin} {et~al.}(2000){Kleeorin}, {Moss}, {Rogachevskii}, \&
  {Sokoloff}}]{Kleeorin+00}
{Kleeorin} N., {Moss} D., {Rogachevskii} I., {Sokoloff} D., 2000, \aap, 361, L5

\bibitem[{{Kleeorin} {et~al.}(2002){Kleeorin}, {Moss}, {Rogachevskii}, \&
  {Sokoloff}}]{Kleeorin+02}
---, 2002, \aap, 387, 453

\bibitem[{{Kleeorin} {et~al.}(1995){Kleeorin}, {Rogachevskii}, \&
  {Ruzmaikin}}]{Kleeorin+95}
{Kleeorin} N., {Rogachevskii} I., {Ruzmaikin} A., 1995, \aap, 297, 159

\bibitem[{{Kleeorin} \& {Ruzmaikin}(1982)}]{Kleeorin+Ruzmaikin82}
{Kleeorin} N., {Ruzmaikin} A.~A., 1982, Magnetohydrodynamics, 18, 116

\bibitem[{{Krause} \& {R\"adler}(1980)}]{Krause+Radler80}
{Krause} F., {R\"adler} K.-H., 1980, {Mean-field {M}agnetohydrodynamics and
  {D}ynamo {T}heory}. Pergamon Press, Oxford

\bibitem[{{Kulsrud} \& {Zweibel}(2008)}]{Kulsrud+Zweibel08}
{Kulsrud} R.~M., {Zweibel} E.~G., 2008, Rep on Prog Phys, 71, 046901

\bibitem[{{Mitra} {et~al.}(2010){Mitra}, {Candelaresi}, {Chatterjee},
  {Tavakol}, \& {Brandenburg}}]{Mitra+10}
{Mitra} D., {Candelaresi} S., {Chatterjee} P., {Tavakol} R., {Brandenburg} A.,
  2010, Astron Nachr, 331, 130

\bibitem[{{Moffatt}(1978)}]{Moffatt78}
{Moffatt} H.~K., 1978, {Magnetic {F}ield {G}eneration in {E}lectrically
  {C}onducting {F}luids}. Cambridge Univ Press, Cambridge

\bibitem[{{Moss}(1995)}]{Moss95}
{Moss} D., 1995, \mnras, 275, 191

\bibitem[{{Moss}(1998)}]{Moss98}
---, 1998, \mnras, 297, 860

\bibitem[{{Moss} {et~al.}(2013){Moss}, {Beck}, {Sokoloff}, {Stepanov},
  {Krause}, \& {Arshakian}}]{Moss+13}
{Moss} D., {Beck} R., {Sokoloff} D., {Stepanov} R., {Krause} M., {Arshakian}
  T.~G., 2013, \aap, 556, A147

\bibitem[{{Moss} {et~al.}(1999){Moss}, {Shukurov}, \& {Sokoloff}}]{Moss+99}
{Moss} D., {Shukurov} A., {Sokoloff} D., 1999, \aap, 343, 120

\bibitem[{{Moss} {et~al.}(2000){Moss}, {Shukurov}, \& {Sokoloff}}]{Moss+00}
---, 2000, \aap, 358, 1142

\bibitem[{{Phillips}(2001)}]{Phillips01}
{Phillips} A., 2001, Geophys Astrophys Fluid Dyn, 94, 135

\bibitem[{{Pouquet} {et~al.}(1976){Pouquet}, {Frisch}, \&
  {Leorat}}]{Pouquet+76}
{Pouquet} A., {Frisch} U., {Leorat} J., 1976, Journal of Fluid Mechanics, 77,
  321

\bibitem[{{R{\"a}dler} {et~al.}(2003){R{\"a}dler}, {Kleeorin}, \&
  {Rogachevskii}}]{Radler+03}
{R{\"a}dler} K.-H., {Kleeorin} N., {Rogachevskii} I., 2003, Geophys Astrophys
  Fluid Dyn, 97, 249

\bibitem[{{Rohde} {et~al.}(1999){Rohde}, {Beck}, \& {Elstner}}]{Rohde+99}
{Rohde} R., {Beck} R., {Elstner} D., 1999, \aap, 350, 423

\bibitem[{Ruzmaikin {et~al.}(1988)Ruzmaikin, Shukurov, \&
  Sokoloff}]{Ruzmaikin+88}
Ruzmaikin A.~A., Shukurov A.~M., Sokoloff D.~D., 1988, Magnetic {F}ields of
  {G}alaxies. Kluwer, Dordrecht

\bibitem[{{Ruzmaikin} {et~al.}(1979){Ruzmaikin}, {Turchaninov}, {Zeldovich}, \&
  {Sokoloff}}]{Ruzmaikin+79}
{Ruzmaikin} A.~A., {Turchaninov} V.~I., {Zeldovich} I.~B., {Sokoloff} D.~D.,
  1979, \apss, 66, 369

\bibitem[{{Shukurov}(1998)}]{Shukurov98}
{Shukurov} A., 1998, \mnras, 299, L21

\bibitem[{{Shukurov}(2005)}]{Shukurov05}
---, 2005, in Lecture Notes in Physics, Berlin Springer Verlag, Vol. 664,
  Cosmic Magnetic Fields, {Wielebinski} R., {Beck} R., eds., p. 113

\bibitem[{{Shukurov}(2007)}]{Shukurov04}
---, 2007, Mathematical Aspects of Natural Dynamos, {Dormy} E., {Soward} A.~M.,
  eds., Chapman \& Hall/CRC, London, pp. 313--359

\bibitem[{Shukurov \& Sokoloff(2008)}]{Shukurov+Sokoloff08}
Shukurov A., Sokoloff D., 2008, in Les Houches, Vol.~88, Dynamos, Cardin P.,
  Cugliandolo L.~F., eds., Elsevier, pp. 251--299

\bibitem[{{Shukurov} {et~al.}(2006){Shukurov}, {Sokoloff}, {Subramanian}, \&
  {Brandenburg}}]{Shukurov+06}
{Shukurov} A., {Sokoloff} D., {Subramanian} K., {Brandenburg} A., 2006, \aap,
  448, L33

\bibitem[{{Subramanian} \& {Brandenburg}(2006)}]{Subramanian+Brandenburg06}
{Subramanian} K., {Brandenburg} A., 2006, \apjl, 648, L71

\bibitem[{{Subramanian} \& {Mestel}(1993)}]{Subramanian+Mestel93}
{Subramanian} K., {Mestel} L., 1993, \mnras, 265, 649

\bibitem[{{Sur} {et~al.}(2007){Sur}, {Shukurov}, \& {Subramanian}}]{Sur+07}
{Sur} S., {Shukurov} A., {Subramanian} K., 2007, \mnras, 377, 874

\bibitem[{{Tabatabaei} {et~al.}(2008){Tabatabaei}, {Krause}, {Fletcher}, \&
  {Beck}}]{Tabatabaei+08}
{Tabatabaei} F.~S., {Krause} M., {Fletcher} A., {Beck} R., 2008, \aap, 490,
  1005

\bibitem[{{Vainshtein} \& {Cattaneo}(1992)}]{Vainshtein+Cattaneo92}
{Vainshtein} S.~I., {Cattaneo} F., 1992, \apj, 393, 165

\bibitem[{{Vishniac} \& {Cho}(2001)}]{Vishniac+Cho01}
{Vishniac} E.~T., {Cho} J., 2001, \apj, 550, 752

\bibitem[{{Vishniac} \& {Shapovalov}(2014)}]{Vishniac+Shapovalov14}
{Vishniac} E.~T., {Shapovalov} D., 2014, \apj, 780, 144

\end{thebibliography}

\label{lastpage}
\end{document}